\documentclass[twocolumn]{aastex631}

\usepackage{xspace}
\usepackage{upgreek}
\usepackage{xcolor}

\definecolor{fullpurple}{RGB}{141,113,212}

\newcommand{\POSEIDON}{\texttt{POSEIDON}\xspace}
\newcommand{\PyMultiNest}{\texttt{PyMultiNest}\xspace}
\usepackage[version=4]{mhchem} 

\newcommand\jwst{\textit{JWST}\xspace}
\newcommand\hst{\textit{HST}\xspace}
\newcommand\spitzer{\textit{Spitzer}\xspace}

\newcommand\teq{T$_\mathrm{eq}$}
\DeclareSymbolFont{UPM}{U}{eur}{m}{n}
\DeclareMathSymbol{\umu}{0}{UPM}{"16}
\newcommand\micro{$\umu$}
\newcommand\microns{\micro m\xspace}

\newcommand{\correction}[1]{\textcolor{black}{#1}}

\begin{document}

\title{The Importance of Optical Wavelength Data on Atmospheric Retrievals \\ of Exoplanet Transmission Spectra}

\correspondingauthor{Charlotte Fairman}
\email{charlotte.fairman@bristol.ac.uk}

\author[0000-0001-9665-5260]{Charlotte Fairman}
\affiliation{HH Wills Physics Laboratory, University of Bristol, Tyndall Avenue, Bristol, BS8 1TL, UK}

\author[0000-0003-4328-3867]{Hannah R. Wakeford}
\affiliation{HH Wills Physics Laboratory, University of Bristol, Tyndall Avenue, Bristol, BS8 1TL, UK}

\author[0000-0003-4816-3469]{Ryan J. MacDonald}
\altaffiliation{NASA Sagan Fellow}
\affiliation{Department of Astronomy, University of Michigan, 1085 S. University Ave., Ann Arbor, MI 48109, USA}

\begin{abstract}
Exoplanet transmission spectra provide rich information about the chemical composition, clouds and temperature structure of exoplanet atmospheres. 
Most exoplanet transmission spectra only span infrared wavelengths ($\gtrsim$\,1\,\microns), which can preclude crucial atmospheric information from shorter wavelengths.
Here, we explore how retrieved atmospheric parameters from exoplanet transmission spectra change with the addition of optical data. From a sample of 14 giant planets with transit spectra from 0.3--4.5\,\microns, primarily from the \textit{Hubble} and \spitzer space telescopes, we apply a free chemistry retrieval to planetary spectra for wavelength ranges of 0.3--4.5\,\microns, 0.6--4.5\,\microns, and 1.1--4.5\,\microns. We analyze the posterior distributions of these retrievals and perform an information content analysis, finding wavelengths below 0.6\,\microns are necessary to constrain cloud scattering slope parameters ($\log{a}$ and $\gamma$) and alkali species Na and K. There is limited improvement in the constraints on the remaining atmospheric parameters. Across the population, we find limb temperatures are retrieved colder than planetary equilibrium temperatures but have an overall good agreement with Global Circulation Models. As \jwst extends to a minimum wavelength of 0.6\,\microns, we demonstrate that exploration into complementing \jwst observations with optical \hst data is important to further our understanding of aerosol properties and alkali abundances in exoplanet atmospheres.
\end{abstract}


\section{Introduction} \label{sec:intro}
Over the past two decades, significant advances have been made in characterizing exoplanet atmospheric properties through transits \citep[e.g.,][]{Charbonneau2002,Kreidberg2015, Sing2016, Wakeford2018,Alderson2023JWSTERS,Xue2023_HD209JWST}, secondary eclipses \citep[e.g][]{Deming2005, Todorov2014, Evans2018, coulombe2023broadband} and full phase curves \citep[e.g.,][]{Knutson2012, Stevenson2017, Mikal2022,Kempton2023_GJ1214}.

Space-based transmission spectroscopy provides a powerful window into exoplanet atmospheres \citep[e.g.,][]{Seager2000,Brown2001}. 
Low resolution spectra can be obtained from transit light curves of telescopes such as \textit{Hubble} (\hst), \spitzer and, \jwst. A majority of early data have been collected from the optical (0.3--1\,\microns) to near-infrared ($>$1\,\microns) using \hst's Space Telescope Imaging Spectrograph (STIS) and Wide Field Camera 3 (WFC3) instruments \citep[e.g.,][]{deming2013ApJ...774...95D,Kreidberg2015, Sing2016, Spake2021}. Photometric observations from \spitzer's IRAC instrument provide additional data points at the infrared end of the spectrum near 3.6 and 4.5\,\microns \citep[e.g.,][]{knutson2011spitzer}. \jwst now fills the gap between 0.6--5\,\microns, providing spectroscopic measurements across the whole near-infrared \citep[e.g.,][]{Alderson2023JWSTERS,ahrer2023JWSTERS,feinstein2023JWSTERS,rustamkulov2023JWSTERS,Xue2023_HD209JWST,Fournier2023MNRAS}, and access to the mid-infrared beyond 5\,\microns with the MIRI instrument \citep[e.g.,][]{grant2023,Dyrek2023}. 

The shape of exoplanet transmission spectra is driven by a planet's chemical composition, cloud composition and temperature structure, which in turn are influenced by the underlying physical processes in an exoplanet atmosphere. Transmission spectra often show an optical scattering slope. This feature can be produced by different physical processes. Rayleigh scattering of photons, by particles smaller than the wavelength of the incident radiation, produces a characteristic $\lambda^{-4}$ slope, whereas scattering from aerosol species can produce steeper gradients \citep{Lecavelier2008}. The optical slope can also be influenced by stellar activity \citep{Pont2013, Barstow2017, Rackham2018}, where the relative number of starspots shifts the transit depth, with this effect being largest towards the shortest wavelengths. High temperatures can also lead to steep optical slopes due to the inflation of the scale height of the atmosphere \citep{Barstow2020} or absorption of high temperature UV species \citep{Lothringer2022}. 

Absorption due to the presence of aerosols/clouds in planetary atmospheres causes the atmosphere to become opaque at pressures greater than the cloud top pressure and thus limits the depth to which we can probe an atmosphere. If clouds are present at the pressures accessed by transmission spectra, they will have the effect of removing the baseline of spectral features. This affects the relationship between feature amplitude and species abundance \citep{Barstow2021}. 

Disentangling aerosol properties from gas phase species (e.g., Na, K, \ce{H2O}) has been the focus of a number of population studies due to the implications of \ce{H2O} abundance on the formation and evolution of exoplanets. Exploring this problem, one of the first major comparative studies \citep{Sing2016} examined the spectra of 10 hot Jupiters in transmission. 
To make comparisons between the planets, they defined two indices $\Delta Z_{UB-LM}$ and $\Delta Z_{J-LM}$ (measuring the strength of scattering to molecular absorption and the relative strength between the mid-infrared continuum and mid-infrared molecular absorption respectively) and measure the strength of the characteristic $1.4 \mu m$ \ce{H2O} absorption feature. A correlation between a strong scattering index and muted water absorption was found leading to the conclusion that clouds and hazes are responsible for weak spectral features, rather than an intrinsic low abundance of \ce{H2O}.


A series of atmospheric retrieval studies have been performed on exoplanet transmission spectra to quantify the impact of aerosols on the \ce{H2O} abundance \citep[e.g.,][]{Barstow2017,Pinhas2019,Welbanks2019metallicity} finding that all spectra are likely affected by cloud opacity to some degree but that a large number of planets also have a low intrinsic abundance of \ce{H2O} compared to solar values.
Studies performed just using \hst WFC3 data from 1.1--1.7\microns covering a single \ce{H2O} band also showed that gray clouds are favored over non-gray scattering for retrievals on near-infrared only data\citep[e.g.,][]{FisherHeng2016,Tsiaras2018}.
It is additionally noted in \citet{FisherHeng2016} that the 1.1--1.7\microns wavelength range is insufficient to overcome degeneracies between modelling choices.

A number of studies looking at individual planetary datasets have demonstrated that the omission of observational measurements in the optical can result in major differences on the retrieved information \citep[e.g.,][]{Wakeford2018,Pinhas2019,Alderson2022}. 
However, there does not exist a detailed systematic study on this effect across a population of planets, and you cannot infer this information by combining multiple study results across different model set-ups and data ranges due to the differing assumptions used in each separate analysis. 

The aim of this study is to explore the dependence of spectral range on the retrieved atmospheric parameters of exoplanet transmission spectra, focusing on the impact of data at optical wavelengths. 
We explore 14 exoplanets (see Table~\ref{tab:observations}) using transmission spectra that come from previously published reductions of, primarily, \hst and \spitzer observations which span from 0.3--4.5\,\microns. This is especially pertinent as \jwst only goes down to a minimum of 0.6\,\microns and there is not yet a clear assessment of how crucial this short wavelength information may be on inferred atmospheric properties below this cut-off. We describe our retrieval set-up and testing in Section \ref{sec:model_setup}. We then present our sample of 14 exoplanet spectra and the results of our retrievals in Section \ref{sec:results} with a table of our datasets and planet parameters in Appendix\,\ref{sec:Appendix}. We calculate the information content obtained with increased wavelength coverage in Section \ref{sec:ICcontent}. In Section \ref{subsec:population} we discuss trends across the population in comparison to the planetary equilibrium temperature. We summarize our results and conclusions in Section \ref{sec:conclusions}.

\begin{deluxetable*}{ccc}
    \tablecaption{Transmission spectra sources for our 14 planet optical data investigation.\label{tab:observations}}
    \tablewidth{0pt}
    \tablehead{
    \colhead{Planet} & \colhead{Instruments} & \colhead{Reference}
    }
    \startdata
    HAT-P-1 b & STIS G430/G750, WFC3 G141, Spitzer IRAC 3.6/4.5 & \cite{Sing2016} \\
    HAT-P-11 b & STIS G430/G750, WFC3 G102/G141, Spitzer IRAC 3.6/4.5 & \cite{Chachan2019} \\
    HAT-P-12 b & STIS G430/G750, WFC3 G141, Spitzer IRAC 3.6/4.5 & \cite{Sing2016} \\
    HAT-P-32 b & STIS G430/G750, WFC3 G141, Spitzer IRAC 3.6/4.5 & \cite{Alam2020} \\
    HD189733 b & STIS G430/G750, WFC3 G141, NICMOS photometry, Spitzer IRAC 3.6/4.5 & \cite{Sing2016} \\
    HD209458 b & STIS G430/G750, WFC3 G141, Spitzer IRAC 3.6/4.5 & \cite{Sing2016} \\
    WASP-6 b & VLT FORS2, STIS G430/G750, WFC3 G141, Spitzer IRAC 3.6/4.5 & \cite{Carter2020} \\
    WASP-12 b & STIS G430/G750, Spitzer IRAC 3.6/4.5 & \cite{Sing2016} \\
    & WFC3 G141 & \cite{Kreidberg2015} \\
    WASP-17 b & STIS G430/G750, WFC3 G141, Spitzer IRAC 3.6/4.5 & \cite{Alderson2022}\\
    WASP-19 b & STIS G430/G750, WFC3 G141, Spitzer IRAC 3.6/4.5 & \cite{Sing2016} \\
    WASP-31 b & STIS G430/G750, WFC3 G141, Spitzer IRAC 3.6/4.5 & \cite{Sing2016} \\
    WASP-39 b & VLT FORS2, STIS G430/G750, WFC3 G102/G141, Spitzer IRAC 3.6/4.5 & \cite{Wakeford2018} \\
    WASP-121 b & STIS G430/G750 & \cite{Evans2018} \\
           & WFC3 G141 & \cite{Evans2016} \\
    WASP-127 b & STIS G430/G750, WFC3 G141, Spitzer IRAC 3.6/4.5 & \cite{Spake2021} \\
     \enddata
\tablecomments{The planets considered in this investigation are chosen based on the availability of at least one transit observation in either of the optical STIS bandpasses (G430/G750) and one observation in either of the WFC3 bandpasses (G102/G141).}
\end{deluxetable*}

\section{Designing a robust and consistent retrieval set-up} \label{sec:model_setup}

To investigate the wavelength dependence of retrieved parameters from exoplanet transmission spectra, we use the open-source retrieval code, \POSEIDON \citep{MacDonaldMadhusudhan2017,Macdonald2023}, which couples a forward model and radiative transfer treatment with a Bayesian retrieval framework for parameter estimation and model selection. Underpinning the retrieval framework is the nested sampling package, \PyMultiNest \citep{Buchner2014}. We limit the implementation of \POSEIDON to free-chemistry retrievals for a 1-dimensional atmospheric parameterization appropriate for the precision of the spectral data analyzed. 
To preserve a consistent retrieval framework across the study, we test a range of model initializations on the $0.3$ to $4.5$\,\microns transmission spectrum of HD\,209458b and define a base retrieval setup with an appropriate level of complexity for each of the spectra considered in our study. We then test a range of wavelength cutoffs on the spectra of HD\,209458b and WASP-39b to determine a base study for our 14 exoplanets. 
\correction{The retrieval configurations for our population are described below.}

\subsection{Forward Model and Retrieval Configuration} \label{subsec:retrieval_config}

There are two practical limits to the retrieval complexity that we considered: the number of model parameters must be sufficiently small to perform a statistically valid retrieval, and computational complexity must be minimized such that a population analysis is feasible.
We define our base atmospheric model with 12 free parameters (see Table~\ref{tab:priors table} for the parameters and their prior ranges) and describe below the selection of these parameters and the overall retrieval set-up.
We first test the different parameterisations available in POSEIDON for the pressure-temperature (P--T) structure of the atmosphere. After retrieving a profile consistent with an isothermal atmosphere using a 5-parameter P-T profile \citep{MadhusudhanSeager2009}, we choose to adopt an isothermal profile, where a uniform temperature prior is defined between $(0.4-1.15)\ T_\mathrm{eq}$. This reduces the number of free parameters in the fit for the P-T profile from 5 to 1. The reference radius at which the atmosphere has a pressure of 10 bar ($R_\mathrm{P, ref}$) is also allowed to vary as a free parameter. 

We assume a bulk atmosphere of \ce{H2} and \ce{He}, with a fixed He/\ce{H2} ratio of 0.17, and include six chemical opacity sources \ce{Na} and \ce{K} \citep{Barklem2016}, \ce{H2O} \citep{Polyansky2018}, \ce{CO2} \citep{Tashkun2011}, \ce{CO} \citep{Li2015} and \ce{CH4} \citep{Yurchenko2018}.
For each we apply uniform priors on the log$_{10}$ mixing ratios of $-12$ to $-1$ dex. 
The presence of alkali species \ce{Na} and \ce{K} in the optical and \ce{H2O} in the near infrared can be inferred from Hubble STIS and WFC3 observations respectively. Additional opacity from carbon species covers the spectral range probed by the Spitzer photometric points. Opacity from nitrogen species (e.g. \ce{HCN}, \ce{NH3}) are excluded from the base set up to reduce model complexity. 
Additional continuum opacity arising from \ce{H2}-\ce{H2} and \ce{H2}-\ce{He} collision induced absorption \citep{karman2019cia}, and Rayleigh scattering are included within the model setup. 
The final four parameters encode cloud and haze opacity into the atmospheric model, using a deck-haze prescription as defined in \cite{MacDonaldMadhusudhan2017}. A cloud deck with infinite opacity across all wavelengths is defined at a pressure level $\log{P_{cloud}}$ and coupled with a wavelength dependent scattering haze. Two parameters describe the behavior of the haze: the scattering slope $\gamma$, and $\log{a}$, that acts as a scaling factor to the \ce{H2} Rayleigh scattering cross Section defined at 350\,nm. To model inhomogeneous cloud cover, $\bar{\phi}_{clouds}$ parameterizes the terminator cloud fraction. We tested models with and without  $\bar{\phi}_{clouds}$ and found that it was necessary to improve the statistical fit to the data in a range of cases and therefore include it in our base retrieval. 

We initialize the atmospheric model on a pressure grid of 100 layers, uniform in log-pressure, ranging from 100--$10^{-9}$ bar. The pressure minimum is selected to allow for the haze parameterisation to fit spectra with strong optical scattering slopes, where the shortest wavelengths probe the lowest atmospheric pressures in transmission. 
To decrease computation time, opacities are sampled from a high-resolution line-by-line database onto a user defined resolution grid. A range of opacity grid resolutions were tested and $R$\,=\,5,000 was selected, as this provided the same level of consistency across the wavelengths covered in this study as an $R$\,=\,10,000 model while reducing the computation time.  
Each retrieval was computed with $N$\,=\,4,000 live points to effectively explore the prior phase space.


\begin{deluxetable}{ccc} \label{tab:priors table}
\tablecaption{Priors for our \POSEIDON retrievals.}
\tablewidth{0pt}
\tablehead{
\colhead{Parameter} & \colhead{Prior Distribution} & \colhead{Prior range}
}
\startdata
R$_{p, \, \mathrm{ref}}$ & uniform & $(0.85 - 1.15)\ R_{p}$ \\
T$_p$ & uniform & $(0.4 - 1.15)\ T_\mathrm{eq}$ \\
log H$_2$O & uniform & $-12$ -- $-1$ \\
log CO$_2$ & uniform & $-12$ -- $-1$ \\
log CO & uniform & $-12$ -- $-1$ \\
log CH$_4$ & uniform & $-12$ -- $-1$ \\
log Na & uniform & $-12$ -- $-1$ \\
log K & uniform & $-12$ -- $-1$ \\
$\log{a}$ & uniform & $-4$ -- $8$ \\
$\gamma$ & uniform & $-20$ -- $2$ \\
$\log{P_\mathrm{cloud}}$ & uniform & $-6$ -- $3$\ dex \\
$\phi_\mathrm{clouds}$ & uniform & $0 - 1$ \\
\tablebreak
\hline
\multicolumn{3}{c}{High Temperature Species} \\
\hline
log TiO & uniform & $-12$ -- $-1$ \\
log VO & uniform & $-12$ -- $-1$ \\
\tablebreak
\hline
\multicolumn{3}{c}{Stellar Parameters} \\
\hline
$f_\mathrm{het}$ & uniform  & $0$ -- $0.5$ \\ 
$T_\mathrm{het}$ & uniform & $(0.6 - 1.2)\ T_\mathrm{eff}$\\
$T_\mathrm{phot}$ & Gaussian & $\mu = T_\mathrm{eff}, \ \sigma = 100 \ K$ \\
\enddata
\tablecomments{The first 12 parameters define the base retrieval configuration. The reference radius is defined at 10\,bar.  $R_{p}$ is the planetary white-light radius. The high-temperature species (TiO and VO) are added for planets with \teq $>$ 2000\,K, while the stellar heterogeneity parameters are included for known active stars.}
\end{deluxetable}

\newpage 

\subsection{Optical Data Cutoffs: Case Studies for HD~209458~b and WASP-39b }\label{sec:wavelength tests}

We first consider the impact of optical data wavelength cutoffs on retrievals of HD 209458b and WASP-39b.
We do not consider measurements from JWST in this study to enable a consistent retrieval and analysis across a larger set of planets. 
We sequentially cutoff short wavelength data points that fall below a threshold $\lambda_\mathrm{min}$, in intervals of 0.1\,\microns, from 0.3 to 1.1\,\microns. Due to the absence of WFC3 G102 data for HD\,209458b, no retrieval is run for $\lambda_\mathrm{min}$\,=\,1.0\,\microns for this planet as there is no data change from the surrounding cut-offs. 

\begin{figure*}
    \centering
    \includegraphics[width=\textwidth]{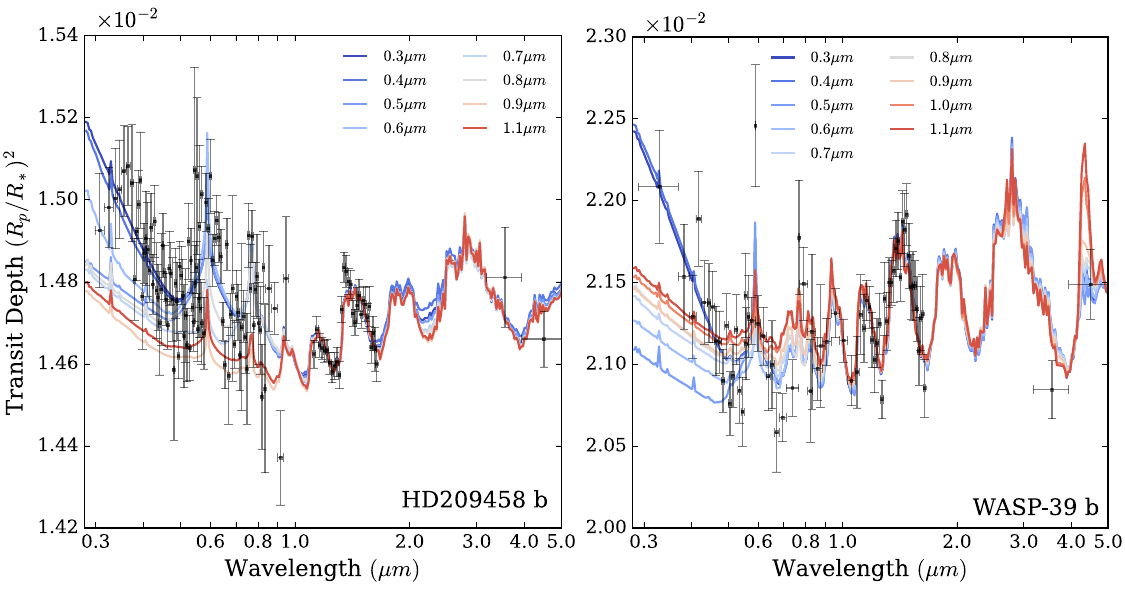}
    \caption{Median retrieved spectra for the \hst and \spitzer observations of HD\,209458b (left) and WASP-39b (right) a lower wavelength data cutoff of $\lambda_\mathrm{min}$. Shown are $\lambda_\mathrm{min}$ values from 0.3\,\microns (dark blue) to 1.1\,\microns (dark red) in intervals of 0.1\,\microns. The observed, multi-instrument transmission spectrum is overlaid (black diamonds). The retrieved spectrum is plotted at R = 100 for clarity.}
    \label{fig:best_fit_retrieved_spectra}
\end{figure*}

Figure \ref{fig:best_fit_retrieved_spectra} shows the median retrieved spectra for each cut-off range combined with the observed data for HD\,209458b (left) and WASP-39b (right). 
Increasing wavelength coverage does not lead to a convergence between retrieved spectra until data is included below 0.5\,\microns. Without data below this point the retrieved median spectra are not well fit at optical wavelengths, where models are unconstrained by observations. 
For WASP-39b, the decrease in optical transit depths with increasing wavelength coverage is driven by the downturn in datapoints preceding the sodium line. Wavelengths below 0.5\,\microns are necessary to fit the observed scattering slope.

\begin{figure*}
    \centering
    \includegraphics[width=0.80\textwidth]{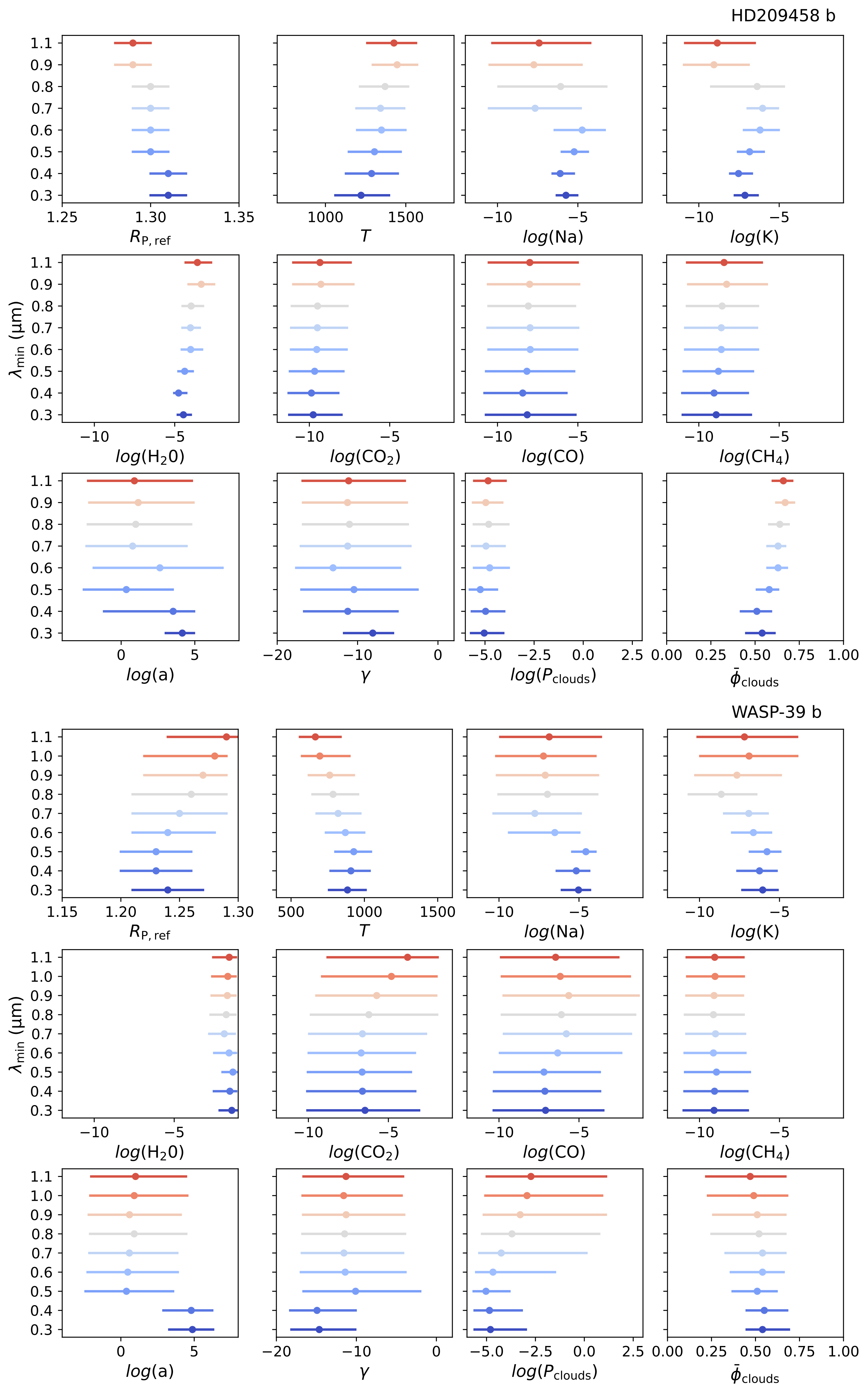}
    \caption{Retrieved median model parameters and $1\sigma$ errors for HD\,209458b (top) and WASP-39b (bottom) from observations spanning a wavelength range $\lambda_\mathrm{min}$ to 4.5\,\microns. $\lambda_\mathrm{min}$ takes values from 0.3 (dark blue) to 1.1\,\microns (dark red) in intervals of 0.1\,\microns.}
    \label{fig:cutoff_parameters}
\end{figure*}

Cloud opacity and alkali species are the main parameters driving the differences between retrieved spectra. A comparison of the median retrieved parameters and their 1$\sigma$ errors for each wavelength range of HD\,209458b and WASP-39b can be seen in Figure~\ref{fig:cutoff_parameters}.  For both planets, \ce{K} abundances are constrained for data below 0.7\,\microns corresponding to spectra where the wavelength range encompasses the line peak at $\sim$\,0.770\,\microns. The characteristic \ce{Na} doublet is found at at $\sim$\,0.589\,\microns. WASP-39b shows constraints for sodium abundance below 0.5\,\microns, however for HD\,209458b, a minimum wavelength of 0.6\,\microns is sufficient to improve constraints on sodium due to the evidence of strong line broadening around the \ce{Na I} feature seen in the spectrum.

Of the four cloud parameters, for HD\,209458b the retrievals across all wavelength ranges can converge on a solution for the cloud top pressure, $\log{P_{cloud}}$, and patchy cloud, $\phi_{clouds}$, parameters. The variation in the median patchy cloud parameter is what drives the variation in the base transit depths between the \ce{H2O} and \ce{CH4} features in the 2--3\,\microns wavelength range. In contrast, the scattering slope parameters $\log{a}$ and $\gamma$ are only well constrained in the full 0.3 to 4.5\,\microns retrieval. For WASP-39b, constraints for $\log{P_{cloud}}$ and $\bar{\phi}_{clouds}$ consistently improve with increased short wavelength coverage. $\log{a}$ and $\gamma$ require wavelengths shortward of 0.5\,\microns for retrievals to constrain the parameter space. 

Whilst the \ce{H2O} constraints become tighter when shorter wavelength information is included in retrievals, the median \ce{H2O} abundance of HD\,209458b for the $\lambda_\mathrm{min}$\,=\,0.3\,\microns retrieval ($-4.46^{+0.46}_{-0.35}$ dex) only marginally falls outside of the $1 \sigma$ errors of the $\lambda_\mathrm{min}$\,=\,1.1\,\microns retrieval. The WFC3 G141 datapoints alone can constrain the water abundance to $-3.59^{+0.85}_{-0.73}$ dex despite the large range of potential cloud solutions. Water is retrieved at the upper boundary of the prior for WASP-39b across all wavelength ranges. This is consistent with the \ce{H2O} abundance found by \cite{Wakeford2018}.
Those species with opacities only accessed by the two Spitzer photometric points (\ce{CO}, \ce{CO2}, \ce{CH4}) remain unconstrained with consistent distributions across all retrieved wavelengths, due to the sparse data. 

\section{Population Analysis}\label{sec:results}

We now extend our investigation on the wavelength dependence of retrieved atmospheric parameters to a total sample of 14 planets: HAT-P-1b, HAT-P-11b, HAT-P-12b, HAT-P-32b, HD\,189733b, WASP-6b, WASP-12b, WASP-17b, WASP-19b, WASP-31b, WASP-121b and WASP-127b, additional to HD\,209458b and WASP-39b. The transmission spectra used come primarily from previously published reductions of \hst and \spitzer observations, with data originating from five instrument modes: \hst STIS/G430L, \hst STIS/G750L, \hst WFC3/G102, \hst WFC3/G141, and the \spitzer IRAC photometric bandpasses centered at 3.6 and 4.5\,\microns. We also consider additional ground-based observations from VLT/FORS2 for WASP-39b and WASP-6b, as reduced by \citet{Wakeford2018} and \citet{Carter2020}, and include the \hst NICMOS photometric points for HD\,189733b that were also used by \citet{Sing2016} and \citet{Barstow2017} in their analysis. Our full sample and data sources are listed in Table~\ref{tab:observations}.

Our planet selection was based on several factors. First, the planet must have transit observation in the optical STIS bandpasses (G430/G750) and one observation in either of the WFC3 bandpasses (G102/G141). While a number of planets in our study are being observed with JWST for consistency in analysis methods we choose not to include those datasets here. Second, we select planets with a range of \teq spanning a range of predicted cloud properties \citep[e.g.,][]{Gao2020,ohno2020ApJ}.
Finally, we limit our sample to the 14 shown due to limitations in computation time for this study.  The planetary and stellar properties for each system are summarized in Appendix~\ref{sec:app:planetary parameters}, Table~\ref{tab:planet parameters}.

We explore multiple optical wavelength cutoffs throughout our planet sample. From our initial findings for HD\,209458b and WASP-39b (Section~\ref{sec:wavelength tests}), we select three cutoffs: $\lambda_\mathrm{min}$\,=\,0.3, 0.6 and 1.1\,\microns. The 1.1\,\microns cutoff marks the transition between \hst WFC3 G141 and G102 observations and hence when using this cutoff the main opacity sources are the 1.15\,\microns and 1.4\,\microns water features. We select 0.6\,\microns as the next cutoff. Not only does this mark the minimum wavelength probed by a majority of \jwst instrument modes, but with this cutoff there can be a partial indication of the pressure-broadened \ce{Na} resonance doublet for relatively clear atmospheres. Our final cutoff, $\lambda_\mathrm{min}$\,=\,0.3\,\microns, considers the full spectral range of the \hst STIS data. 

\subsection{Population Retrieval Configuration} \label{subsec:pop_retrieval_config}

We consider three different retrieval configurations for our population analysis. First, we apply the same base retrieval configuration as in Section~\ref{subsec:retrieval_config} to all planets with $T_{\mathrm{eq}} <$ 2,000\,K. For those planets exceeding 2,000\,K (WASP-12b, WASP-19b, and WASP-121b), we additionally considered a `high-temperature' configuration including \ce{TiO} and \ce{VO} opacity.
Finally, for planets with \emph{a priori} known active host stars (HAT-P-11b, HD\,189733b, WASP-6b, and WASP-19b) we additionally consider a retrieval model accounting for a single population of unocculted stellar heterogeneities to asses any evidence of stellar contamination on the transmission spectrum \citep[e.g.,][]{Rackham2017,pinhas2018}. 
For the two planets where some datasets were not consistently reduced (WASP-121b and WASP-12b), we initially fitted for a relative offset between these data for a preliminary retrieval on the full $\lambda_\mathrm{min}$\,=\,0.3\,\microns wavelength range, with offsets applied to the combined \hst WFC3 and \spitzer data. Our final retrieval results adopt a fixed offset equal to the median value of the retrieved offset. 
We note that due to low spectral resolution, resulting in a small number of measured datapoints, WASP-12b is only analyzed for the 0.6 and 0.3\,\microns cutoffs and WASP-19b is limited to the full 0.3\,\microns cut-off analysis, such that the degrees of freedom in the model does not exceed the number of datapoints. 

Table~\ref{tab:retrieval_config_sample} summarizes which retrieval configurations were conducted for each planetary dataset, with the final selected set-up highlighted. All but two of our spectra could be fit by our base retrieval model, with HAT-P-11b requiring additional stellar contamination parameters and WASP-121b favoring the inclusion of high-temperature species. With our optimal retrieval configurations established for each planet, we proceed to discuss the results of our population analysis. 

\begin{deluxetable}{cccc} \label{tab:retrieval_config_sample}
    \tablecaption{Retrieval configuration applied to each planet.}
    \tablewidth{0pt}
    \tablehead{
    \colhead{Planet} & \colhead{Base} & \colhead{High Temperature} & \colhead{Stellar Contamination}}
    \startdata
    HAT-P-1b & \checkmark \textcolor{cyan}{$\star$} & & \\
    HAT-P-11b & \checkmark && \checkmark \textcolor{cyan}{$\star$} \\
    HAT-P-12b & \checkmark \textcolor{cyan}{$\star$} & \\
    HAT-P-32b & \checkmark \textcolor{cyan}{$\star$} & \\
    HD\,189733b & \checkmark \textcolor{cyan}{$\star$} & & \checkmark \\
    HD\,209458b & \checkmark \textcolor{cyan}{$\star$} & \\
    WASP-6b & \checkmark \textcolor{cyan}{$\star$} & & \checkmark \\
    WASP-12b & \checkmark & \checkmark \textcolor{cyan}{$\star$} & \\
    WASP-17b & \checkmark \textcolor{cyan}{$\star$} & \\
    WASP-19b & \checkmark \textcolor{cyan}{$\star$} & \checkmark $\dagger$ & \checkmark $\dagger$ \\
    WASP-31b & \checkmark \textcolor{cyan}{$\star$} & \\
    WASP-39b & \checkmark \textcolor{cyan}{$\star$} & \\
    WASP-121b & \checkmark & \checkmark \textcolor{cyan}{$\star$} & \\
    WASP-127b & \checkmark \textcolor{cyan}{$\star$} & \\
    \enddata
\tablecomments{Ticks (\checkmark) indicate which setups were tested for each planet, while the stars (\textcolor{cyan}{$\star$}) mark the final adopt retrieval setup for the population analysis (Section~\ref{sec:population analysis}). For WASP-19b, daggers ($\dagger$) mark reduced parameter setups, where the carbon species \ce{CH4}, \ce{CO} and \ce{CO2} are removed from the retrieval. All planets are tested with the base setup.}
\end{deluxetable}

\subsection{Results: Population-level Analysis }\label{sec:population analysis}

\begin{figure*}
    \centering
    \includegraphics[width=0.9\textwidth]{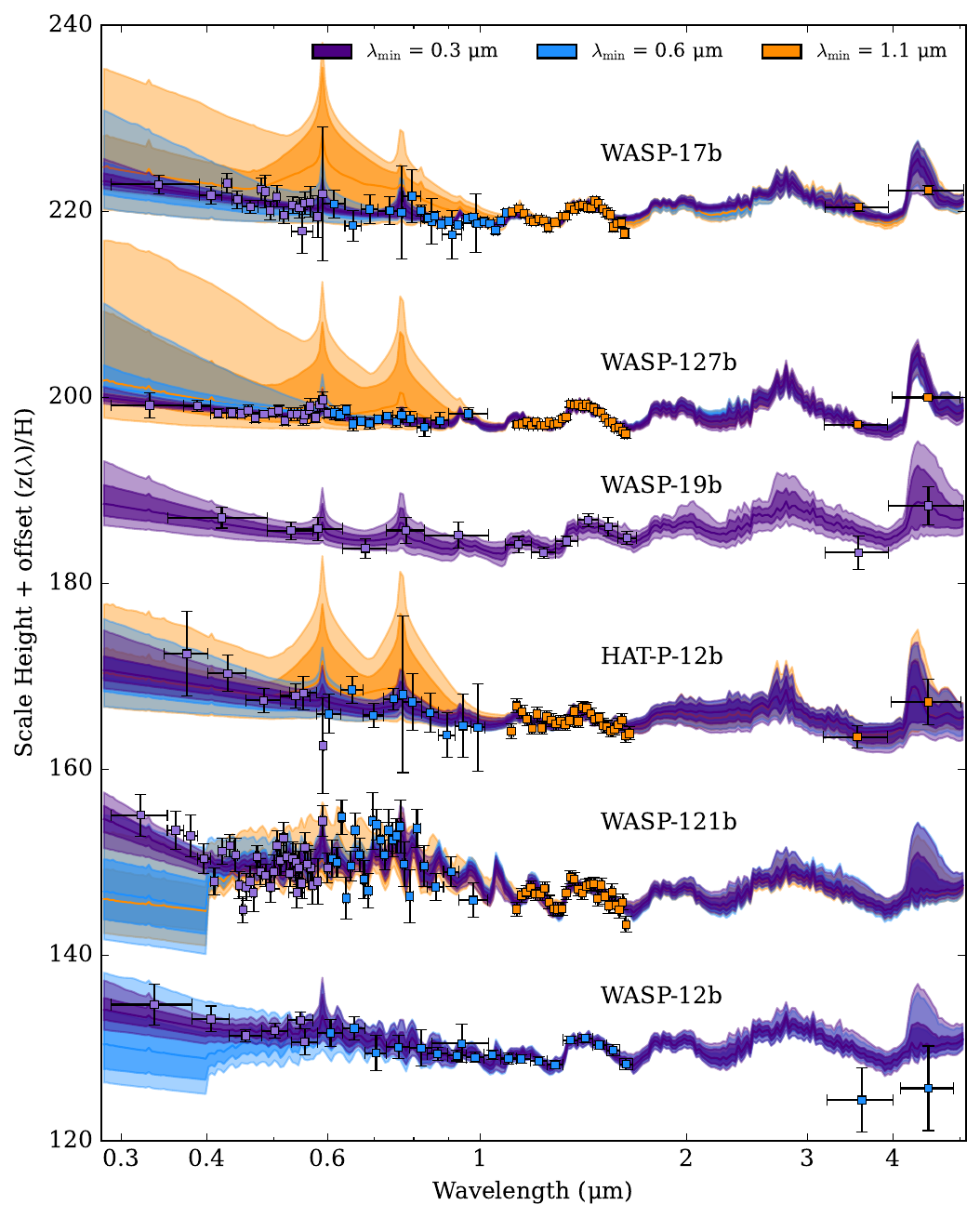}
    \caption{Retrieved transmission spectra for the population. Retrieved median spectra for each wavelength range, $\lambda_\mathrm{min}$\,=\,0.3\,\microns, $\lambda_\mathrm{min}$\,=\,0.6\,\microns and $\lambda_\mathrm{min}$\,=\,1.1\,\microns, are displayed in purple, blue and orange respectively. The light and dark shaded regions correspond to the $1$ and $2\sigma$ confidence regions. The datapoints show the observed spectrum, where the marker colors correspond to data above the minimum wavelength of the three retrievals, following the same color scheme as the retrieved spectra. Planets are displayed in order of their retrieved Rayleigh enhancement factor, $\log{a}$, with the planet with the lowest value of $\log{a}$ at the top of the figure. The y-axis scale is expressed in atmospheric scale heights, $H = k_B T / \mu g$, where $T$ is taken as the retrieved temperature from the $\lambda_\mathrm{min}$\,=\,0.3\,\microns retrieval (+ an arbitrary offset for clarity).}
    \label{fig:spectra1}
\end{figure*}

\begin{figure*}
    \centering
    \includegraphics[width=0.9\textwidth]{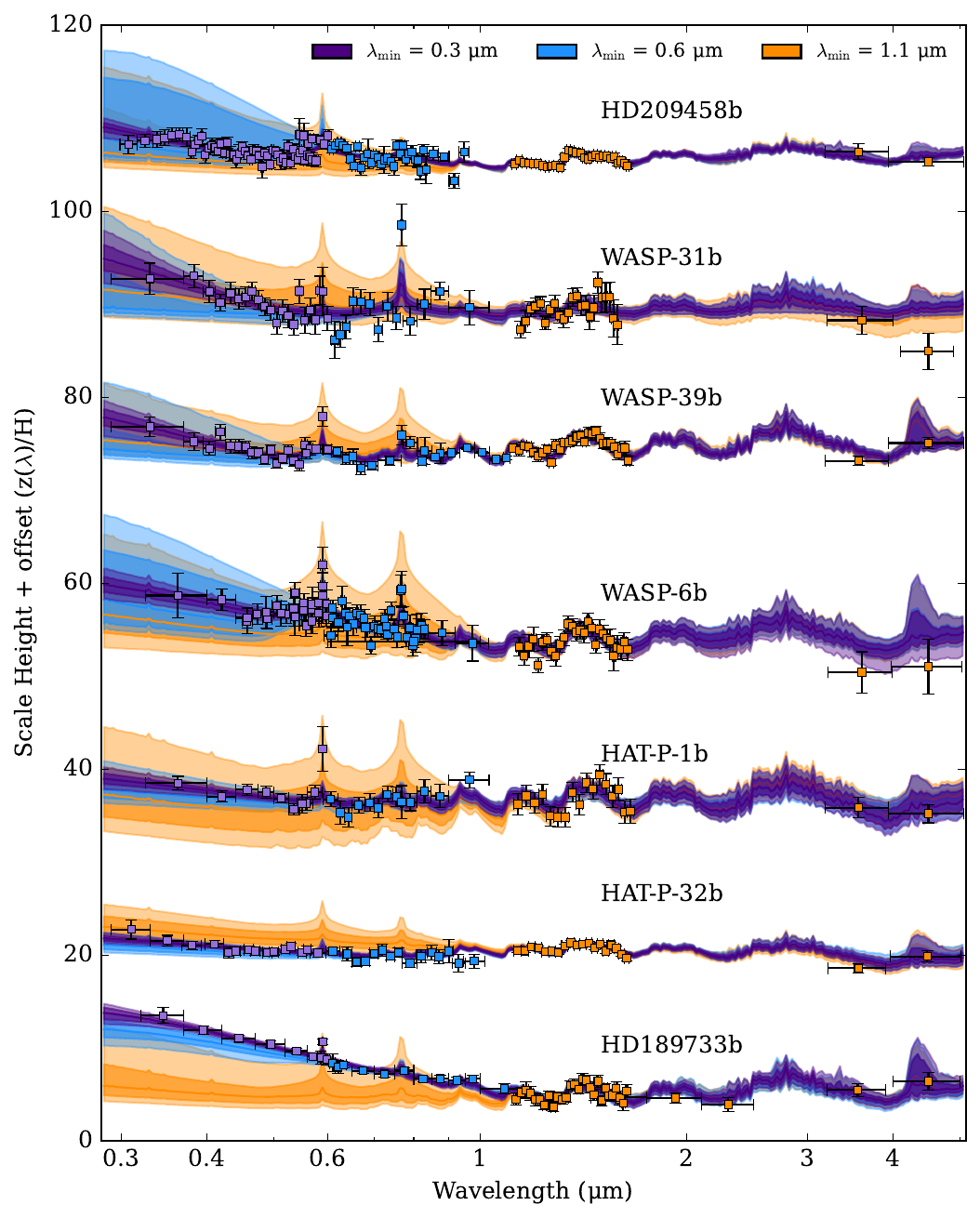}
    \caption{Retrieved median transmission spectra for the population (continued). See Figure~\ref{fig:spectra1} for the caption.}
    \label{fig:spectra2}
\end{figure*}

Figures~\ref{fig:spectra1} and \ref{fig:spectra2} present the retrieved transmission spectra across the population for our three optical data cutoffs. The planets are ordered from the lowest retrieved optical slope Rayleigh enhancement factor (WASP-17b), $\log{a}$, to the highest retrieved value (HD\,189733b). The corresponding retrieved atmospheric properties (posterior median and $1\sigma$ confidence intervals) for each planet and spectral range are shown in Figure~\ref{fig:pop_cutoff_pars}. 
\correction{Full posterior distributions for all planets are provided in the supplementary online material:
\\ \dataset[10.5281/zenodo.10407463]{https://doi.org/10.5281/zenodo.10407463}.}
Below, we discuss the effect of changing the spectral range on model parameters. 

\begin{figure*}
    \centering
    \includegraphics[width=0.9\textwidth]{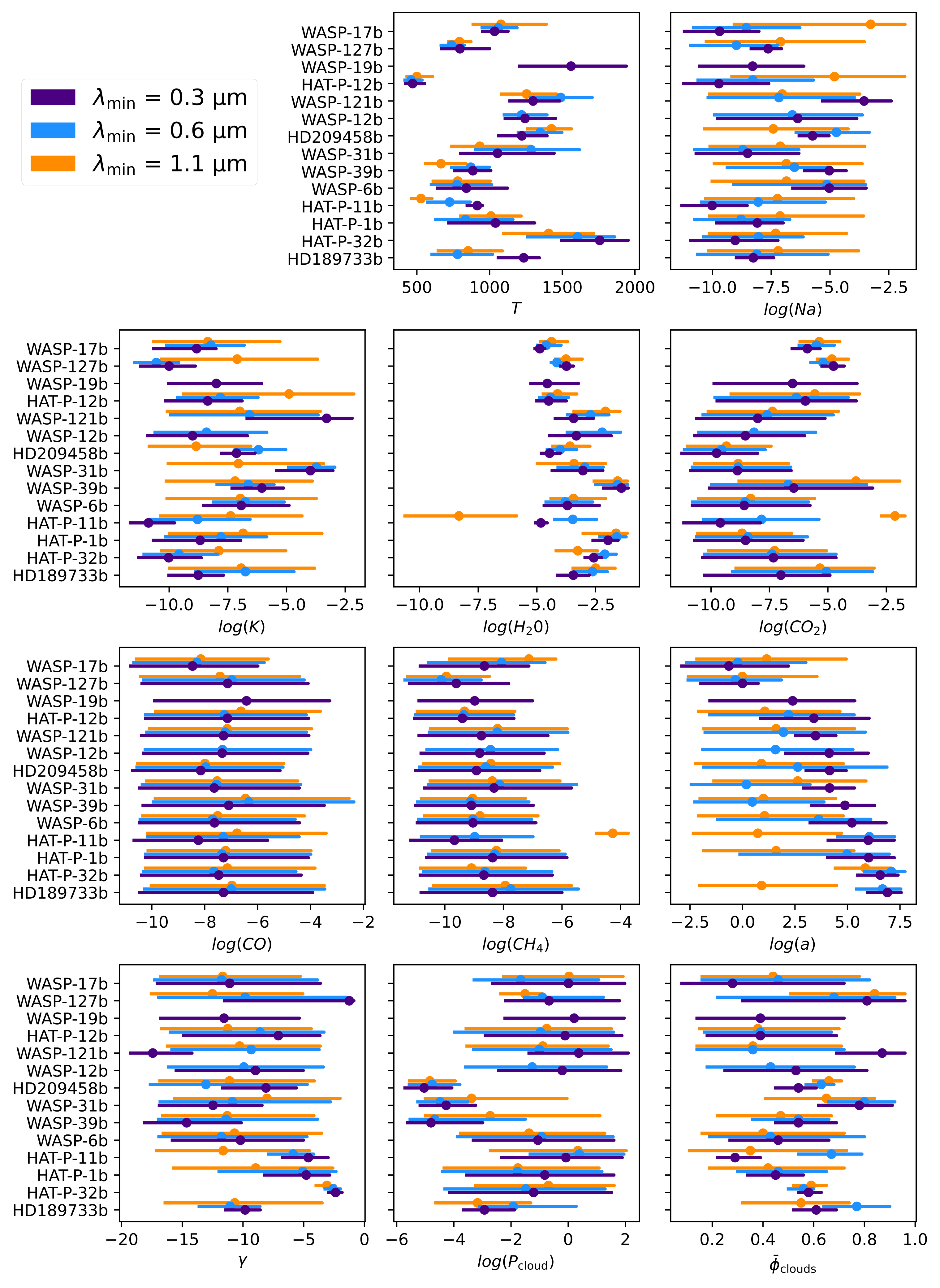}
    \caption{Median retrieved parameters and $1\sigma$ errors across the population. Only parameters from the base retrieval are displayed (HAT-P-11b, WASP-12b and WASP-121b contain additional parameters in their retrievals). The retrieval setup of each planet can be found in Table~\ref{tab:retrieval_config_sample}.  Results for each wavelength range, $\lambda_\mathrm{min}$\,=\,0.3\,\microns, $\lambda_\mathrm{min}$\,=\,0.6\,\microns and $\lambda_\mathrm{min}$\,=\,1.1\,\microns, are displayed in purple, blue and orange respectively. Planets are displayed in order of their retrieved Rayleigh enhancement factor, $\log{a}$, with the planet with the lowest value of $\log{a}$ at the top.}
    \label{fig:pop_cutoff_pars}
\end{figure*}

\subsubsection{Aerosol Properties}

Of the four cloud parameters, $\log{a}$ is the best constrained when using the full wavelength range, $\lambda_\mathrm{min}$\,=\,0.3\,\microns.
We find the average 1$\sigma$ error range for $\log{a}$ decreases from 6.54 to 5.66 to 3.05 for the 1.1, 0.6 and 0.3\,\microns cutoffs, respectively, corresponding to a precision improvement of $46\%$ when adding data from 0.6 to 0.3\,\microns. 
The scattering slope, $\gamma$, also sees improved constraints with increased wavelength coverage, but in several cases it remains largely unconstrained even with the addition of spectral information down to 0.3\,\microns likely due to large uncertainties at these wavelengths. 
For $\gamma$, the error range between the 1.1 and 0.6\,\microns cutoffs show limited improvement, decreasing from 11.7 to 10.5. However, with the addition of 0.3--0.6\,\microns data, the 1$\sigma$ error range decreases by  30$\%$ to 7.2. 
Whilst a significant proportional decrease in the error range, this still leaves large regions of degenerate parameter space between $\log{a}$ and $\gamma$. 
Consequently, it is this shortest wavelength range that is crucial for constraining the scattering slope parameters, where the primary gain in information content comes from the Rayleigh enhancement factor (see Section\,\ref{sec:ICcontent}).

In contrast to the scattering slope parameters, constraints on the cloud-top pressure, $\log{P_{\mathrm{cloud}}}$, do not improve significantly with additional optical wavelength coverage. Even for the full wavelength range down to $\lambda_\mathrm{min}$\,=\,0.3\,\microns, for many planets $\log{P_{\mathrm{cloud}}}$ is largely unconstrained.
Between 1.1 and 0.3\,\microns, the average 1$\sigma$ error is only reduced by 0.57 dex from 4.31 dex, because a majority of spectra result in only an upper limit corresponding to a non-detection of an optically thick cloud-deck. Of the four planets with cloud pressure constraints, HD\,209458b, HD\,189733b, WASP-31b, and WASP-39b, HD\,209458b is the only one constrained at $\lambda_\mathrm{min}$\,=\,1.1\,\microns.  
Because the cloud pressure contributes opacity across the entire spectrum particularly affecting the amplitude of spectral features, in the case of HD\,209458b there is sufficient information from the infrared spectrum to constrain the cloud deck. 
WASP-31b, WASP-39b and HD\,189733b are important counter-examples to this trend, where the addition of optical wavelengths can improve cloud deck constraints if there is evidence of high altitude, gray clouds.

The cloud fraction, $\bar{\phi}_\mathrm{clouds}$, does show some improvement with shorter wavelength coverage. 
The addition of wavelengths between 1.1 and 0.3\,\microns decreases the average 1$\sigma$ error range from 0.45 to 0.32. 
Those planets with well constrained cloud fractions (HD\,189733b, HAT-P-32b, HAT-P-11b) follow a general trend of tighter cloud fractions with increasing $\log{a}$, as can be seen in Figure~\ref{fig:pop_cutoff_pars}. 
Therefore, optical wavelengths can improve our knowledge of the terminator cloud fraction, which must be well-constrained to also precisely measure the temperature and chemical composition of exoplanet atmospheres \citep[e.g.,][]{LineParmentier2016,MacDonaldMadhusudhan2017}. 

Overall, our population analysis shows that increasing the wavelength coverage into the optical can significantly improve constraints on aerosol scattering parameters. 
With only near-infrared observations, these scattering parameters often add unnecessary complexity to the retrieval model. A similar conclusion was drawn by \citet{FisherHeng2016} when retrieving only WFC3 1.1--1.7\,\microns data, where they showed that only gray clouds were required to fit the data. 
Wavelengths shortward of 0.6\,\microns are necessary to obtain constraints on $\log{a}$ and $\gamma$, with additional marginal improvement on $\log{P_{\mathrm{cloud}}}$ and $\bar{\phi}_\mathrm{clouds}$.

\subsubsection{Atmospheric Temperature}

Across the population, we do not see an improvement in retrieved temperatures from adding optical data. 
This average 1$\sigma$ temperature range between the  $\lambda_\mathrm{min}$\,=\,1.1 and 0.3\,\microns spectra decreases from 363 K to 344 K, with the error range increasing marginally for the $\lambda_\mathrm{min}$\,=\,0.63\,\microns to 373 K.
\correction{As the \ce{H2O} feature in the near-infrared is often the most prominent absorption signature at these wavelengths, this feature influences the convergence of the temperature parameter through the atmospheric scale height. }
In principle, Rayleigh scattering can infer temperature from the optical spectrum (for a clear atmosphere), but since we are fitting for a free optical slope this method of determining the temperature is degenerate with the scattering properties \correction{\citep[e.g.,][]{Barstow2020,BarstowHeng2020}}.
The lack of additional temperature information with optical wavelengths is also a consequence of implementing free chemistry retrievals. In equilibrium chemistry retrievals of WASP-39b, \cite{Wakeford2018} find improved temperature constraints with the inclusion of optical data, driven by the opacity of optical species influencing the best fit model temperature. Additionally, degeneracies between the reference pressure and temperature that occur across the population
are generally not broken by the addition of observations at optical wavelengths.

\subsubsection{Chemical Composition}

All three wavelength ranges can broadly constrain the \ce{H2O} abundance, with consistent median values to within an order of magnitude. 
Some improvement in constraints is realized as the wavelength range of the observed spectrum increases, where the average 1$\sigma$ error range in the \ce{H2O} abundance decreases from 2.00, to 1.36 and 1.15 dex, for the $\lambda_\mathrm{min}$\,=\,1.1, 0.6 and 0.3\,\microns spectra respectively. 
In the case of WASP-17b and WASP-127b constrains improve by $>$ 50$\%$. 
For WASP-17b, this improvement is primarily from observations between 0.3 and 0.6 \microns, whereas for WASP-127b, it is the 0.6 to 1.1 \microns wavelength range that contributes to the improved constraints. 
With the exception of HAT-P-11b (see section \ref{sec:hat11}), the retrieved \ce{H2O} abundances are consistent between the three wavelength ranges. 
Primarily, the \ce{H2O} mixing ratio is well characterized by infrared observations shortward of 1.1 \microns where \ce{H2O} molecular bands dominate the opacity.
However, optical data can provide improvements on the \ce{H2O} constraints. 
We may expect the water abundance and constraints to change with increased wavelength coverage through a gain of information on cloud parameters. Studies have discussed the degeneracy between cloud pressure level and water abundance \citep[e.g.,][]{WelbanksMadhusudhan2019} where the 1.4\,\microns water feature can be fit by a lower \ce{H2O} abundance, clear atmosphere or a higher \ce{H2O} abundance atmosphere where the feature is muted by cloud opacity. However, in general, constraints on the cloud pressure level do not significantly improve with the addition of optical wavelengths.

Precise alkali metal abundances are crucially enabled by optical data. With little prominent atomic opacity at long wavelengths, the $\lambda_\mathrm{min}$\,=\,1.1\,\microns spectra are unable to constrain the abundances of \ce{Na} and \ce{K}.  
A minimum wavelength of $0.6$\,\microns is sufficient to constrain the \ce{K} abundance (should its atomic line feature be present). 
For planets with detections of \ce{K} ($>$ 3$\sigma$), the 1$\sigma$ constraints span a consistent range of 2.4 dex for both the $\lambda_\mathrm{min}$ = 0.6 and 0.3\,\microns spectra.
However, to reliably constrain \ce{Na}, the full optical wavelength range is required. For planets with \ce{Na} detection significances $>$ 3$\sigma$, the abundance constraint ranges improve from 4.66 dex to 1.94 dex between the $\lambda_\mathrm{min}$\,=\,0.6 and 0.3\,\microns spectra. 
We find that, with the exception of HD\,209458b, detecting only the red wing of the \ce{Na} resonance feature with \hst (i.e. $\lambda_\mathrm{min}$ = 0.6) is not sufficient to constrain the \ce{Na} abundance.

The \hst and \spitzer data examined here cannot generally constrain the abundances of the carbon-bearing species, nor do the constraints improve with additional wavelength information in the optical. This is demonstrated from the average $1\sigma$ abundance constraints between each cutoff. For \ce{CO}, the range changes negligibly from 6.16 to 5.99 dex between the 1.1 and 0.3 micron cutoffs and for \ce{CO2} and \ce{CH4}, the average constraints change from 4.12 to 4.35 dex and 3.72 to 4.12 dex respectively. 
However, WASP-17b and WASP-127b are outliers in this trend. Across all wavelengths, we are able to constrain the \ce{CO2} mixing ratio to within $<$ 2 dex. 
These constraints are driven by the strong relative offset between the Spitzer points where the 4.5 micron point is at higher transit depths than the 3.6 micron point. This can be fit by a sharp CO2 peak \correction{near} 4.5\,\microns.
Increasing the wavelength coverage into the optical decreases constraints by $\sim$ 30$\%$ between the $\lambda_\mathrm{min}$\,=\,1.1 and 0.3\,\microns spectra.
Therefore, if infrared data can provide evidence of carbon species, the inclusion of optical wavelengths can strengthen constraints on the retrieved mixing ratios. 
\begin{deluxetable}{lccc} \label{tab:detections}
    \tablecaption{Detection significances for \ce{Na}, \ce{K} and \ce{H2O}.}
    \tablewidth{0pt}
    \tablehead{
    \colhead{Planet} & \colhead{\(DS_{Na}\)} & \colhead{\(DS_{K}\)} & \colhead{\(DS_{H_2O}\)}
    }
    \startdata
    HAT-P-1 b & $1.7\sigma$ & $1.3 \sigma$ & $3.7 \sigma$ \\
    HAT-P-11 b & --- & --- & $4.7\sigma$ \\
    HAT-P-12 b & --- & $1.3\sigma$ & $4.6\sigma$ \\
    HAT-P-32 b & --- & $1.1\sigma$ & $8.5\sigma$ \\
    HD\,189733 b & $5.2\sigma$ &  $2.0\sigma$ & $5.7\sigma$ \\
    HD\,209458 b & $7.3\sigma$ & $5.6\sigma$ & $8.9\sigma$ \\
    WASP-6 b & $4.5\sigma$ & $3.5\sigma$ & $4.8\sigma$ \\
    WASP-12 b & $2.2\sigma$ & --- & $6.1\sigma$ \\
    WASP-17 b & --- & --- & $8.4\sigma$ \\
    WASP-19 b & --- & $1.5\sigma$ & $3.2\sigma$ \\
    WASP-31 b & --- & $3.2\sigma$ & $2.6\sigma$ \\
    WASP-39 b & $4.5\sigma$ & $3.2\sigma$ & $8.9\sigma$ \\
    WASP-121 b & $2.4\sigma$ & $2.3\sigma$ & $6.1\sigma$ \\ 
    WASP-127 b & $2.9\sigma$ & $1.1\sigma$ & $14.4\sigma$ \\
    \enddata
    \tablecomments{The equivalent sigma values are computed from Bayesian evidence ratios. `---' indicates a non-detection (Bayes factor $<$ 1).}
\end{deluxetable}

\subsection{Assessment of Individual Planet Spectra}

We next discuss retrieval results for the individual planets that are not captured within the population plots.
\correction{Detection significances for \ce{Na}, \ce{K} and \ce{H2O} are computed from Bayesian evidence ratios, with significances $>$ 1$\sigma$ presented in Table~\ref{tab:detections}.
}

\correction{The posterior distributions for individual planets can be fond in the supplementary online material: \dataset[10.5281/zenodo.10407463]{https://doi.org/10.5281/zenodo.10407463}, to which we direct the reader. 
We note a few general insights to the relationships between retrieved parameters across the planetary population. Widely present across the population and wavelength cutoffs is a correlation between the reference pressure and temperature. With the exception of WASP-17b, which is mentioned below,  this does not improve with the inclusion of optical data. }

\correction{Considering just the $\lambda_{min}$ = 1.1 \microns retrievals, there are many cases where the parameter space is largely degenerate. With the inclusion of optical data, correlations between parameters can emerge as regions of the parameter space are ruled out. Cases where correlations arise with the inclusion of optical data are often between cloud parameters, optical species mixing ratios or across these two parameter categories. For example, we see for HAT-P-1b that the inclusion of optical data reveals a correlation between $\gamma$ and $\bar{\phi}_{clouds}$ while still representing a degeneracy between those two parameters. Further emergent correlations for individual planets are presented in their respective sections.}


\subsubsection{HAT-P-1b}

HAT-P-1b shows increasing improvement on cloud scattering properties, $\log{a}$ and $\gamma$, with increasing wavelength coverage. However we see little improvement in other fit parameters. \ce{H2O} is well constrained with a detection significance of 3.7$\sigma$ but \ce{Na} remains tentative in our analysis despite a significant ``by-eye'' deviation at the \ce{Na I} wavelength.

\subsubsection{HAT-P-12b}

For the $\lambda_\mathrm{min}$\,=\,1.1\,\microns retrieval HAT-P-12b favors high Na and K abundance modes. This condition
is resolved with the addition of observations below 1.1 \microns, where \ce{Na} and \ce{K} features can be measured, which rule out high abundance solutions. Large transit depth uncertainties on the STIS data prevent any strong constraints on the cloud parameters, where broad regions of the parameter space remain degenerate. 
However, the  $\lambda_\mathrm{min}$\,=\,0.3\,\microns retrieval finds, although unconstrained, a $\gamma$ distribution skewed towards values consistent with Rayleigh scattering. This supports the sub-micron particle size ranges found by \citet{Wong2020} from retrievals implementing Mie particle scattering.

\subsubsection{HAT-P-32b}

HAT-P-32b is the only planet in our sample that is consistent between all three wavelength cutoffs. 
Small differences in the \ce{H2O} abundance and temperature are seen between $\lambda_\mathrm{min}$\,=\,1.1\,\microns and 0.6\,\microns, resulting from the non-detection of  \ce{K} and \ce{Na}. However, the scattering slope measured in the G141 bandpass is maintained throughout the optical wavelengths with no significant change in gradient, unlike that seen in WASP-31b or WASP-39b where there is a clear turning point in the gradient of the slope in the optical wavelengths. This means the water feature alone can constrain $\log{a}$ and $\gamma$ to within 1.1 dex.

\subsubsection{HD\,189733b}

HD\,189733b shows the highest retrieved value of the Rayleigh enhancement factor, $\log{a}$, with this parameter constrained even by the $\lambda_\mathrm{min}$\,=\,0.6\,\microns retrieval. 
This is due to the turning point of the prominent scattering slope extending longwards of 0.6\,\microns. This scattering has been previously attributed to stellar activity \citep{McCullough2014}, however, our retrievals fitting for stellar inhomogeneties were not statistically favored over the base retrieval performed.   

For most of the planets, our retrievals are unable to constrain aerosol parameters with $\lambda_\mathrm{min}$\,=0.6\,\microns, since the data from 0.6 and 1.1\,\microns typically contains a large scatter with a wide range of possible aerosol parameters that could fit the spectrum. 
For example, HD\,209458b and WASP-6b also contain a moderate scattering slope that is only constrained with the $\lambda_\mathrm{min}$\,=\,0.3\,\microns spectrum, however, our retrievals cannot disentangle the cloud parameters at $\lambda_\mathrm{min}$\,=\,0.6\,\microns due to the presence of strong \ce{Na} absorption.

\subsubsection{HD 209458b}

HD 209458b is discussed in detail in Section~\ref{sec:wavelength tests}. 
\correction{Emergent correlations between cloud parameters are seen between $\log{a}$ and $\log{P_{cloud}}$ for $\lambda_{min}$ = 0.3 \microns retrievals and the mixing ratios of \ce{H2O}, \ce{Na} and \ce{K} show correlations in both the $\lambda_{min}$ = 0.3 and 0.6 \microns retrievals.}

\subsubsection{WASP-6b}

WASP-6b has significant detections of \ce{Na}, \ce{K}, and \ce{H2O}. The \ce{Na} line wing is detected in $\lambda_\mathrm{min}$\,=\,0.6 \,\microns but is not constrained until $\lambda_\mathrm{min}$\,=\,0.3\,\microns. The significant scattering slope, $\log{a}\,\sim\,$6 and $\gamma$\,=\,-10 is well defined but unconstrained until the full dataset is used. The inclusion of \hst/WFC3 data also allows us to place a near solar constraint on the water abundance, also shown by \citet{Carter2020}. \correction{Correlations between cloud parameters arise at wavelengths below 0.6 \microns, most notably $\log{a}$ -- $\gamma$ and T -- $\gamma$. The \ce{H2O}--\ce{Na}--\ce{K} correlation present for HD\,209458b can also be seen in the $\lambda_{min}$ = 0.3 \microns posterior distribution for WASP-6b.}

\subsubsection{WASP-12b}

The \hst data for WASP-12b from \citet{Sing2016} only permits loose constraints on cloud properties. With only six datapoints shortward of 0.6\,\microns, no inferences can be made about the cloud fraction across the terminator and only the lowest cloud deck pressures can be ruled out. 

\subsubsection{WASP-17b}

WASP-17b has poorly constrained aerosol parameters due to the low-precision STIS data (caused by a partial transit, as detailed by \citealt{Alderson2022}). For all wavelength ranges cloud constraints do not improve, contrary to the general trend across the population. However, inclusion of wavelengths below 0.6\,\microns improves constraints on the $R_{p, \, \mathrm{ref}}$--$T$ degeneracy. WASP-17b is the only planet where this degeneracy is largely broken, as seen in the reference pressure posterior distribution. 

However, the short wavelength data does improve the chemical abundance constraints for WASP-17b. Most notably, correlations between the \ce{H2O}, \ce{K}, and \ce{CO2} abundances collapse with the addition of $<$ 0.6\,\microns data. Our infrared-only retrieval ($\lambda_\mathrm{min}$\,=\,1.1\,\microns) finds a high abundance mode for \ce{Na} and \ce{K} (see Figure~\ref{fig:best_fit_retrieved_spectra}), causing the median retrieved spectrum to show significantly greater transit depths around the \ce{Na} and \ce{K} features than for the $\lambda_\mathrm{min} =$ 0.6 and 0.3\,\micron retrievals. 
This high abundance mode corresponds to a set of solutions that fit the data with a high mean molecular weight and high temperature atmosphere. However, this unphysical mode is lost for WASP-17b with the additional information from the alkali wings at shorter wavelengths. This suggests that the near-infrared \hst WASP-17b data can inaccurately and over-confidently predict alkali abundances.

\subsubsection{WASP-19b}

Our WASP-19b retrieval is limited due to the small number of datapoints in this planet's spectrum, however, our results broadly agree with previous studies with retrieved parameters consistent with those in \citet{Pinhas2019}, despite their study using a more complex 19-parameter retrieval. 
We note that some retrieval studies have used different data for WASP-19b. In particular, \citet{Welbanks2019metallicity} use VLT/FORS2 data from \citet{Sedaghati2017} for their analysis, leading them to find constraints on \ce{Na} and \ce{TiO}. We choose not to include the \citet{Sedaghati2017} data in our analysis, as it is reduced with a significantly different analysis method to that of the \citet{Sing2016} dataset. Including the VLT/FORS2 data would require additional free offset parameters in our retrievals, as well as the consideration of wavelength-correlated data, which would run counter to our uniform analysis approach. However, we did test retrievals with and without stellar activity for WASP-19b \citep[see e.g.,][]{Espinoza2019,Sedaghati2021} and found that stellar heterogeneity is not statistically favored by the \hst data.

\subsection{WASP-31b}

WASP-31b provides a clear demonstration of optical data resolving ambiguous aerosol properties. With only near-infrared data ($\lambda_\mathrm{min}$\,=\,1.1\,\microns), there is
a bimodal solution in WASP-31b's optical slope between Rayleigh and super-Rayleigh scattering solutions, with the most likely mode being a Rayleigh scattering slope. The addition of data down to 0.6\,\microns reduces the likelihood of this mode and leads to a flatter, more unconstrained posterior distribution. However, once the data down to 0.3\,\microns are included the super-Rayleigh region of parameter space favored. 

We note that other chemical species, beyond those considered here, may also be present in WASP-31b's atmosphere. \citet{MacDonald2017b} reported tentative evidence of NH$_3$ in WASP-31b's WFC3 data, while  \citet{Braam2021} proposed the presence of CrH given the STIS observations. Recently, \citet{Flagg2023} confirmed CrH via high-resolution Doppler spectroscopy. Further, our 3.2$\sigma$ detection of \ce{K} in WASP-31b is driven by a single data point at 0.77\,\microns, which may be due to instrument systematics. The validity of this detection has been debated, since ground-based observations have been unable to reproduce the detection \citep[e.g.,][]{Mcgruder2020, Gibson2019, Gibson2017}

\subsubsection{WASP-39b}

WASP-39b is discussed in detail in Section~\ref{sec:wavelength tests}.\\

\subsubsection{WASP-121b}

WASP-121b is one of our high-temperature case studies, for which TiO and VO opacity must be included in the retrieval analysis. The multiple strong peaks between 0.4\,\microns and 1.0\,\microns for the $\lambda_\mathrm{min}$\,=\,0.6 and 1.1\,\microns retrievals indicate prominent contributions from \ce{TiO} and \ce{VO} opacity. 
Without optical data below 0.6\,\microns, TiO and VO, alongside the aerosol parameters, are unconstrained. \correction{Additionally, WASP-121b shows emergent correlations between mixing ratios of \ce{VO} and \ce{H2O} with $\log{a}$ when observations below 0.6 \microns are included. For all wavelength ranges, a strong correlation between \ce{VO} and \ce{H2O} is present, and constraints do not improve with the addition of shorter wavelengths.}

Our retrievals find a relatively steep slope is needed to fit the optical spectrum, however, this may result from our model setup. Such an optical slope may be mimicking opacity from other high-temperature species beyond \ce{TiO} and \ce{VO} that we do not include in our model. We note that both \citet{Evans2016} and \citet{Evans2018}, from which we obtain our spectra, attribute the large transit depth shortwards of 0.4\,\microns to other opacity sources, without the need for a scattering slope. Therefore, absorbers such as \ce{HS} and \ce{FeH} may provide an alternative explanation for WASP-121b's high transit depth at short optical wavelengths. 

\subsubsection{WASP-127b}

WASP-127b is one of a handful of planets where the temperature constraints worsen with increased optical wavelength coverage (from spanning 184 K for the $\lambda_\mathrm{min}$\,=\,0.3\,\microns spectrum to 333 K for the $\lambda_\mathrm{min}$\,=\,0.3\,\microns spectrum). For the $\lambda_\mathrm{min}$\,=\,0.3\,\microns retrieval, a region of higher temperature parameter space, which is disfavored by the $\lambda_\mathrm{min}$\,=\,0.6 and 1.1\,\microns retrievals, occupies a portion of the probability distribution. This high-temperature region only arises for sub-Rayleigh scattering values of $\log{a}$ and $\gamma$, which are most prominent for the 0.3\,\microns retrieval and completely absent from the 1.1\,\microns retrieval. This example demonstrated that shallow slopes in the optical data can alter other retrieval atmospheric properties from the values inferred from the infrared alone.

\subsubsection{HAT-P-11b}\label{sec:hat11}

HAT-P-11b is an outlier within our planet sample, occupying the Neptune regime of mass-radius space. 
It orbits a known active K star with a measured spot covering fraction of $3^{+6}_{-1}\%$ \citep{Morris2017THEONE!}.
Despite correcting for spot coverage in the data reduction \citep{Chachan2019}, HAT-P-11b is the only planet in our sample where retrievals support the inclusion of stellar contamination. 
Therefore, we can use HAT-P-11b to provide a test on the wavelength dependence of retrieved parameters for a stellar contaminated transmission spectrum.

When only considering near-infrared data, we find spurious evidence of \ce{CH4} and \ce{CO2} in HAT-P-11b's atmosphere. As seen in Figure~\ref{fig:pop_cutoff_pars}, the mixing ratios of \ce{CH4} and \ce{CO2} are bounded and constrained to high values for $\lambda_{\mathrm{min}} = 1.1$\,\microns. With an equilibrium temperature of 840\,K, HAT-P-11b lies on the boundary between a \ce{CO} and \ce{CH4} dominated atmosphere, depending on the metallicity of the atmosphere \citep{Moses2013}, so the presence of \ce{CH4} and \ce{CO2} within the atmosphere is physically plausible. However, once the short wavelength data is included these detections disappear. 

\begin{figure}
    \centering
    \includegraphics[width=\columnwidth]{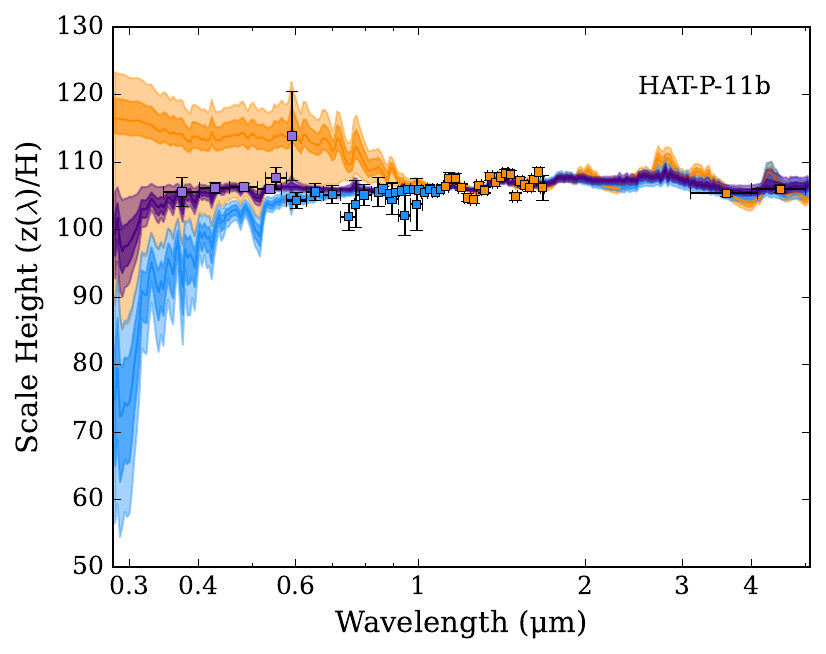}
    \caption{Retrieved transmission spectrum for HAT-P-11b. Retrieved median spectra for each wavelength range, $\lambda_\mathrm{min}$\,=\,0.3\,\microns, $\lambda_\mathrm{min}$\,=\,0.6\,\microns and $\lambda_\mathrm{min}$\,=\,1.1\,\microns, are displayed in purple, blue and orange respectively. Light and dark shaded regions correspond to the $1$ and $2\sigma$ bounds. The datapoints show the observed spectrum, where the marker colors correspond to data above the minimum wavelength of the three retrievals, following the same color scheme as the retrieved spectra. The y-axis scale is expressed in atmospheric scale heights, $H = k_B T / \mu g$, where $T$ is taken as the retrieved temperature from the $\lambda_\mathrm{min}$\,=\,0.3\,\microns retrieval.}
    \label{fig:spectra HAT-P-11b}
\end{figure}

Similarly, the retrieved \ce{H2O} abundance is significantly anomalous for the  $\lambda_{\mathrm{min}} = 1.1$\,\microns retrieval. 
Without data between 0.3\,\microns and 1.1\,\microns, the stellar heterogeneity temperature is retrieved approximately 1200\,K below the effective temperature of the star ($T_{\mathrm{eff}}$ = 4780\,K), corresponding to unocculted starspots. The resulting low \ce{H2O} abundance can therefore be attributed to starspots mimicking the atmospheric \ce{H2O} absorption features, which can be seen by the increase in transit depth at optical wavelengths in the  $\lambda_{\mathrm{min}} = 1.1$\,\microns retrieval (see Figure~\ref{fig:spectra HAT-P-11b}).

However, once shorter wavelength optical data is included ($\lambda_{\mathrm{min}}$ = 0.6 and 0.3\,\microns), the retrieved spectra for HAT-P-11b support solutions with stellar heterogeneities hotter than the surrounding photosphere (i.e. unocculted faculae). This produces a strong negative slope in the transmission spectrum, which then no longer biases the \ce{H2O} abundance. Similarly to retrievals without stellar contamination, wavelengths below 0.6\,\microns are necessary for cloud parameter constraints.

Although the inclusion of stellar contamination improves the model fit for HAT-P-11b, the retrieved model still provides a poor fit to the data ($\chi^{2}_{\nu} = 2.42$) and may not accurately encompass the underlying physics of the planet. A potential drawback of the setup is the limitations of the stellar contamination model used here, which only considers a single stellar heterogeneity population with a single temperature. For active stars, this model may struggle to account for the realistic distribution of heterogeneities due to the inability to model both starspots and faculae together.

\begin{figure*}[ht!]
    \centering
    \includegraphics[width=\textwidth]{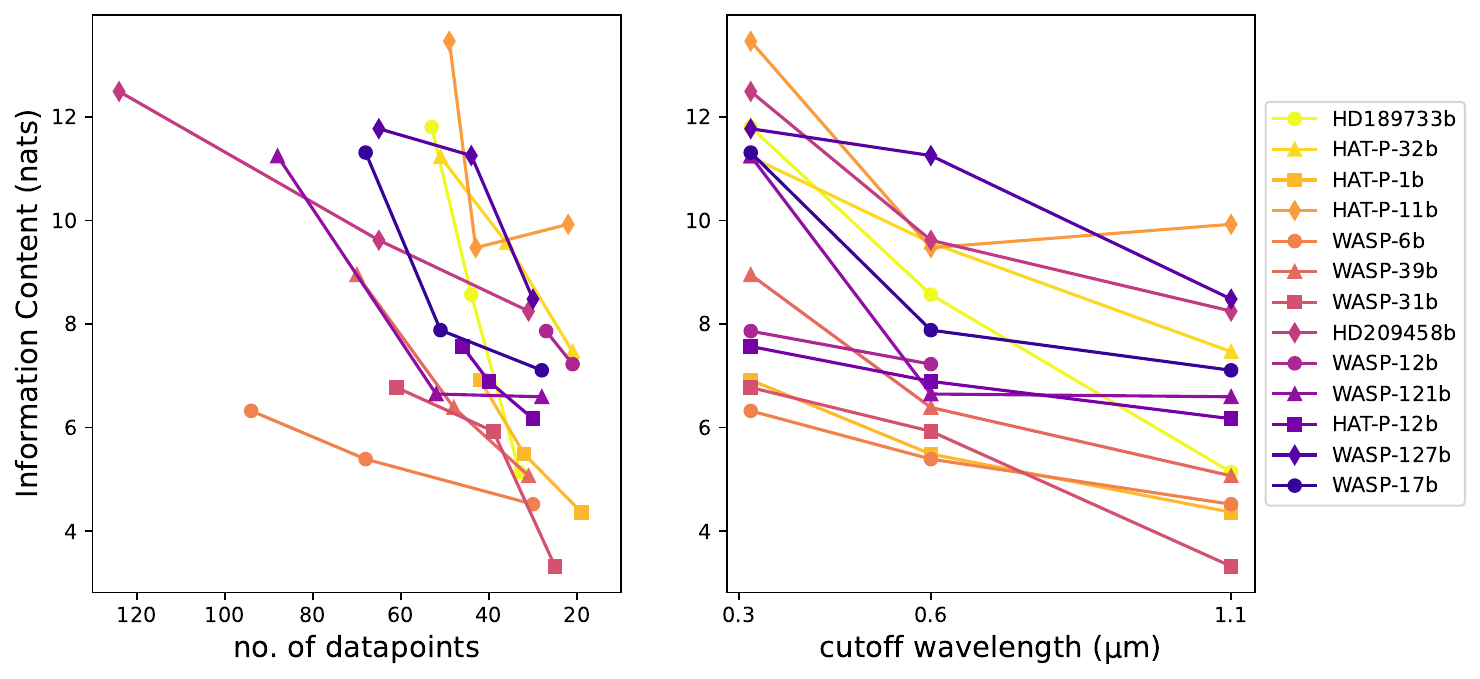}
    \caption{The information content (measured in nats) between the posterior and prior distribution for $\lambda_{min}$ = 0.3, 0.6 and 1.1\microns retrievals. Left panel: the change in information content with the number of datapoints. Right panel: the change in information content with data wavelength cutoff. Planets are ordered in terms of their retrieved Rayleigh enhancement factor, $\log{a}$, with the planet with the highest value of $\log{a}$ (HD\,189733b) in yellow and the lowest value of $\log{a}$ (WASP-17b) in purple.}
    \label{fig:total IC}
\end{figure*}

\section{Information content analysis}\label{sec:ICcontent}

Whilst we can show the importance of the inclusion of optical wavelengths in constraining atmospheric parameters through their retrieved median values and errors, we can go further to quantitatively estimate the gain/loss in information with increased spectral range at short wavelengths.
As this study focuses on observational data, there is a complex relationship between the underlying composition of the planet’s atmosphere, observational uncertainty (through the signal strength, transit depth errors and binning), model parameterization and the information gained through retrievals. 
This makes quantitative conclusions across the population difficult to disentangle. 
However, on a planet-by-planet basis, we can quantify the information gain through entropy estimation. 

Information content (IC) analysis has been applied to inverse problems within exoplanet atmospheric characterization in several previous studies \citep[e.g.,][]{bennekeseager2012information, BatalhaLine2017, howe2017information}. Information content analyses can quantify how the knowledge of an atmospheric state changes relative to the prior, following an observation. 
This knowledge change can be computed from the difference in entropy of the prior and posterior distributions of a retrieval. 

The mutual information, or information content of a retrieval is defined as the change in entropy between the prior and the posterior distribution. In this case, the entropy of a distribution is the average information required to encode a parameter value from its prior.
Therefore, information content describes the improvement in the confidence of the retrieved atmospheric parameters, given a set of observations, from the initial prior knowledge \citep{line2012information}. 
We define the information content ($I(\theta, X)$) of a retrieval, where $\Theta$ is the set of all parameters $\theta$ and $X$ is the set of observations $x$ as

\begin{equation}
    I(\Theta, X) = H(\Theta) - H(\Theta|X)
\end{equation}
where $H(\Theta)$ is the entropy of the prior distribution $p(\Theta)$ and $H(\Theta|X)$ is the entropy of the posterior distribution $p(\Theta|X)$.

For a discrete random variable $X$, entropy is defined as 

\begin{equation}
    H(X) = -\sum_{x \in X} p(x)\log{p(x)} 
\end{equation}
which is the expectation of the self information $h(x) = -\log{p(x)}$. 
Extending the entropy to a continuous distribution function $F$ of a random variable $X$, with an associated probability density function $f(x)$, gives the following definition

\begin{equation}\label{eq:continuous entropy}
    H(f) = -\int^{\infty}_{-\infty} f(x)\log{f(x)} dx
\end{equation}

The output of our retrievals provides posterior samples of some unknown theoretical distribution function. As such, we use the package \texttt{scipy.stats.differential$\_$entropy} \citep{alizadeh2015entropy} to implement the Vasicek method as an entropy estimator of our posterior samples. 
\cite{vasicek1976information} expresses Equation~\ref{eq:continuous entropy} as

\begin{equation}
    H(f)  = \int^1_0 \log{\Bigg\{\frac{d}{dp} F^{-1}(p)\Bigg\}dp}
\end{equation}
which is then reformulated by replacing $F$ with the empirical distribution function $F_n$. The derivative of $F_n$ can be estimated from the ordered samples from the probability distribution $x_n$, where the window size, $m$, must be a positive integer and $m < n/2$. This reformulation, $HV_{mn}$ is Vasikeck's entropy estimator, defined as

\begin{equation}\label{eq:vasicek}
    HV_{mn} = \frac{1}{n} \sum^n_{i=1}\log{\Big\{\frac{n}{2m}(X_{i+m} - X_{i-m})\Big\}}
\end{equation}

The estimator is consistent, such that $HV_{mn} \to H(f)$ as $n \to \infty$, $m \to \infty$, and $m/n \to 0$. 

\begin{figure*}[ht!]
    \centering
    \includegraphics[width=\textwidth]{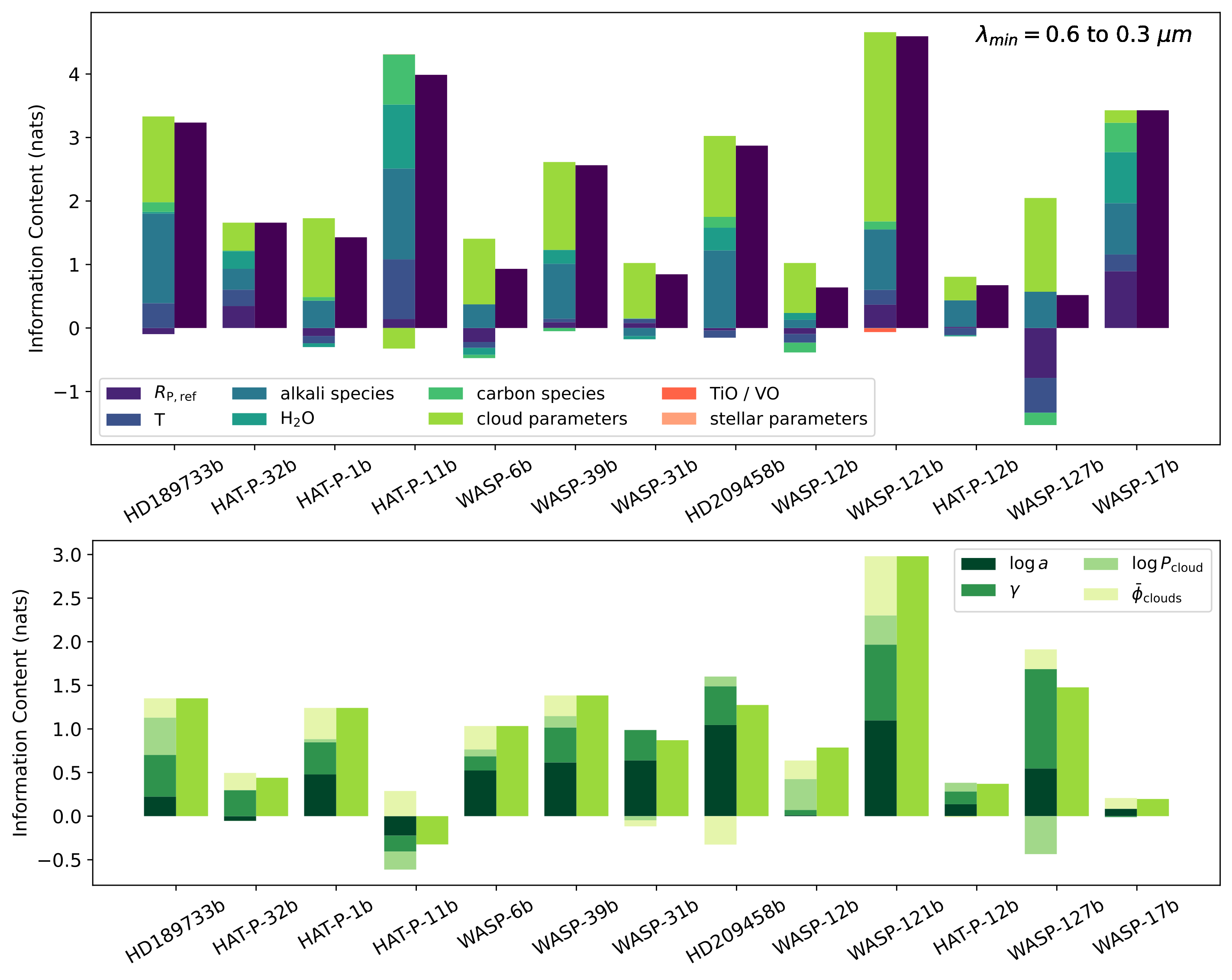}
    \caption{Information content (measured in nats) broken down by parameter group for $\lambda_\mathrm{min}$ = 0.6 to 0.3\microns. Top: breakdown of parameters by $R_{p,\mathrm{ref}}$, T, alkali species, \ce{H2O}, carbon species, cloud parameters, high temperature species (\ce{TiO}/\ce{VO}) and stellar parameters. Bottom: breakdown of the information content change per cloud parameter. Planets are ordered from highest to lowest (left to right) retrieved Rayleigh enhancement factor, $\log{a}$.}
    \label{fig:IC 0.6 0.3 with clouds} 
\end{figure*}

We estimate the information content between the prior and posterior distributions for each model parameter by drawing random samples from the prior distributions and using the output samples from the marginalized posterior distribution. To these samples, we apply Vasicek’s entropy estimator, where we measure information content in nats. For an event with probability $1/e$, the information content in nats is one. We then sum over the mutual entropy for all parameters to find the total information content gained from the retrieval for a given set of observations. We also assess the information content between the posteriors of retrievals with different spectral ranges. 

\begin{figure}
    \includegraphics[width=\columnwidth]{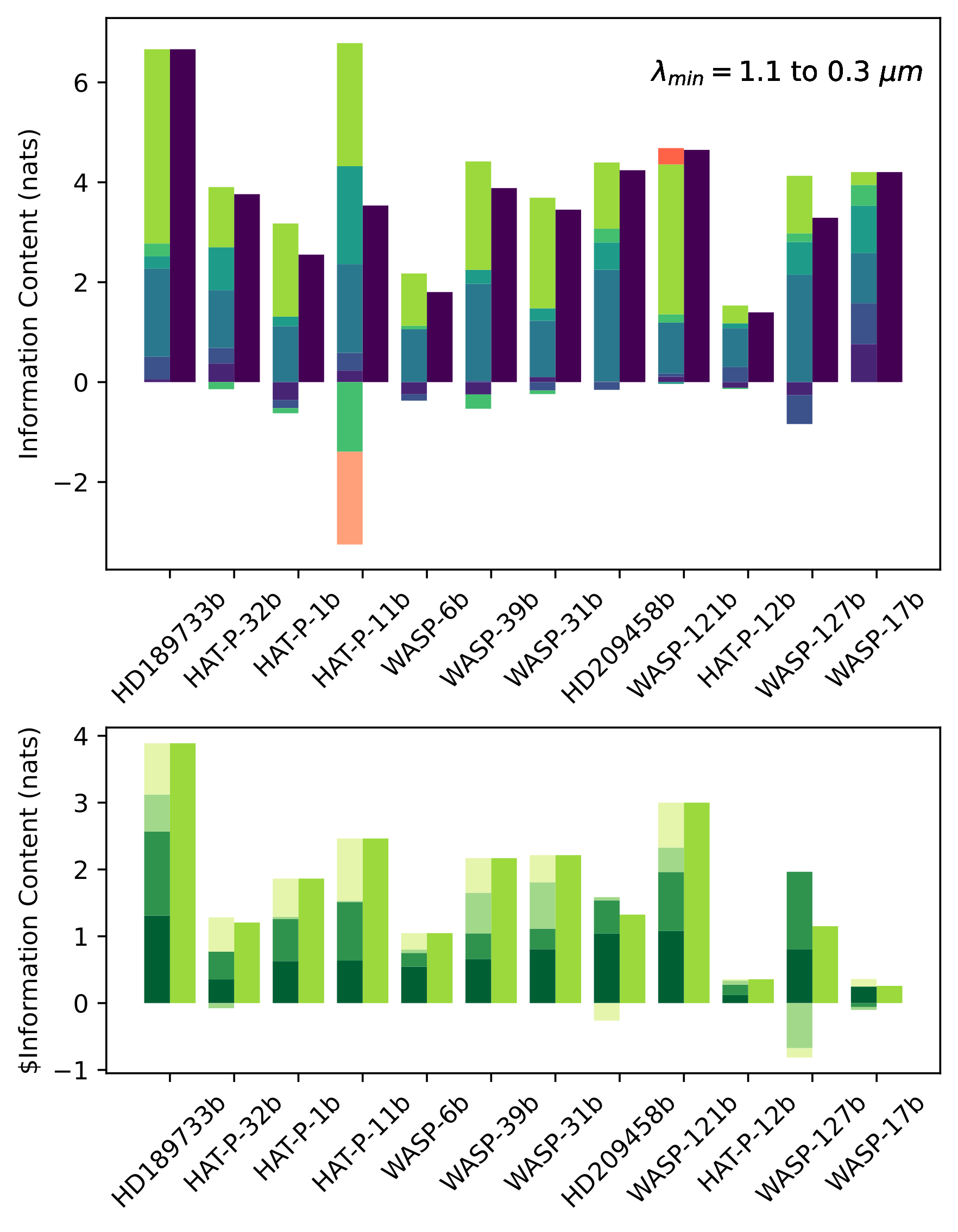}
    \caption{Information content (measured in nats) broken down by parameter group for $\lambda_\mathrm{min}$ = 1.1 to 0.3\microns. Top: breakdown of parameters by $R_{P,\mathrm{ref}}$, T, alkali species, \ce{H2O}, carbon species, cloud parameters, high temperature species (\ce{TiO}/\ce{VO} and stellar parameters. Bottom: breakdown of the information content change per cloud parameter. Planets are ordered from highest to lowest (left to right) retrieved Rayleigh enhancement factor, $\log{a}$.}
    \label{fig:IC 1.1 0.3 with clouds} 
\end{figure}

\subsection{Information Content Analysis Results}

Figure~\ref{fig:total IC} displays the change in information between the prior and posterior distribution from our retrievals. We show the information content change as a function of the number of datapoints (left panel) and with the different optical wavelength cutoffs (right panel) for each planet. As a trend, the IC increases with greater wavelength and datapoint coverage. Most planets have a greater increase in IC per wavelength and per datapoint between $\lambda_\mathrm{min}$\,=\,0.6\,\microns and 0.3\,\microns than between the $\lambda_\mathrm{min}$\,=\,1.1\,\microns to 0.6\,\microns retrievals, with some showing a marked increase in IC with only a few additional datapoints in this range.  Exceptions to this trend are WASP-31b and WASP-127b (and also HAT-P-32b, but only for the datapoint difference, not the spectral range). The average increase in information between 1.1\,\microns and 0.6\,\microns is 1.39 nats and the average increase in information between 0.6\,\microns and 0.3\,\microns is 2.11 nats for the sample of exoplanets. 


We find that the greatest average information gain across the planet population is found between 0.6\,\microns and 0.3\,\microns. Therefore, we explore the contribution of each parameter to the IC in this wavelength range by calculating the change in estimated entropy between the posterior distributions of the $\lambda_\mathrm{min}$ = 0.6\,\microns and 0.3\,\microns retrievals. 
In this case, it is possible for the change in entropy to be negative, as it is calculated between two final measurements and not a prior and posterior distribution.
\correction{Figure~\ref{fig:IC 0.6 0.3 with clouds} shows} the breakdown of the change in IC for each portion of the model between the $\lambda_\mathrm{min}$ = 0.6\,\microns and 0.3\,\microns retrievals. We first show the contribution from the clouds across all four parameters and then show the breakdown in IC for each cloud parameterization used in the model. 

We see that the change in information content does not show the same ordering as the median retrieved Rayleigh enhancement factor ($\log{a}$). 
It can be seen that cloud parameters dominate the IC gain between 0.6\,\microns and 0.3\,\microns. If we further break down the cloud parameters, the greatest contribution to information gain does come from $\log{a}$, followed by $\gamma$. 
It is unsurprising that the scattering parameters provide the greatest information gain, as this supports the finding in Section~\ref{sec:population analysis} that optical wavelengths provide the strongest constraints for scattering parameters. 

Significant contributions to information between 0.6\,\microns and 0.3\,\microns also come from the alkali species. In particular, we find the greatest gains in information for those planets with the highest detection significances of \ce{Na} (see Table~\ref{tab:detections}).
For planets where there is no detection of alkali species (e.g. WASP-17b and HAT-P-32b), there is still a contribution to the information content due to retrievals ruling out high abundances of these species. 
Interestingly, as the wavelength range of retrievals increases, for some planets, we lose information on $R_\mathrm{P, ref}$, temperature, and carbon species. These changes are likely related to the increased wavelength probing a wider pressure range.

WASP-17b does not follow the trend of cloud parameters dominating the increase in information with data over an extended optical wavelength range. 
Instead, the increase in information is driven by improved constraints on \ce{H2O}, alkali species and the reference pressure. 
The lack of cloud information reiterates the findings of Section~\ref{sec:population analysis}, where limited constraints on the cloud parameters are found, due to the large uncertainties of the STIS data. 

HAT-P-11b also acts as an outlier. 
It is the only planet where the information on cloud parameters decreases with the addition of wavelengths below 0.6\,\microns. 
Whilst the IC between 0.6\,\microns and 0.3\,\microns is negligible, between 1.1\,\microns and 0.3\,\microns (Figure~\ref{fig:IC 1.1 0.3 with clouds}) a large decrease in information is seen where the stellar parameters no longer converge on an unocculted cold spot solution.
The stellar contamination information is driven by optical data below 1.1\,\microns. 
However, due to the ability of stellar contamination to mimic atmospheric molecular features, stellar contamination will directly impact the retrieved abundances of species such as \ce{H2O}, which itself impacts the temperature of the atmosphere.
As such, HAT-P-11b  shows a large change in IC across many parameters. 

Finally, Figure~\ref{fig:IC 1.1 0.3 with clouds} displays the same information content breakdown by parameter as Figure~\ref{fig:IC 0.6 0.3 with clouds}, but between the $\lambda_\mathrm{min}$ = 1.1\,\microns and 0.3\,\microns retrievals.
This yields the same result in information content as between 0.6\,\microns to 0.3\,\microns but increases the impact of alkali species as the contribution of \ce{K} opacity between 0.6\,\microns and 1.1\,\microns is added. 
Of the planet population, HD\,189733b has the greatest increase in information content between 1.1\,\microns and 0.3\,\microns but not 0.6\,\microns to 0.3\,\microns which can be explained by the scattering slope extending to wavelengths greater than  0.6\,\microns. 
The scattering parameters $\log{a}$ and $\gamma$ are well constrained for the $\lambda_\mathrm{min}$ = 0.6\,\microns retrieval such that less additional information is gained with the inclusion of wavelengths below 0.6\,\microns.

\section{Cloud and Temperature Trends Across the Population}\label{subsec:population}

\begin{figure}
    \centering
    \includegraphics[width = 0.48\textwidth]{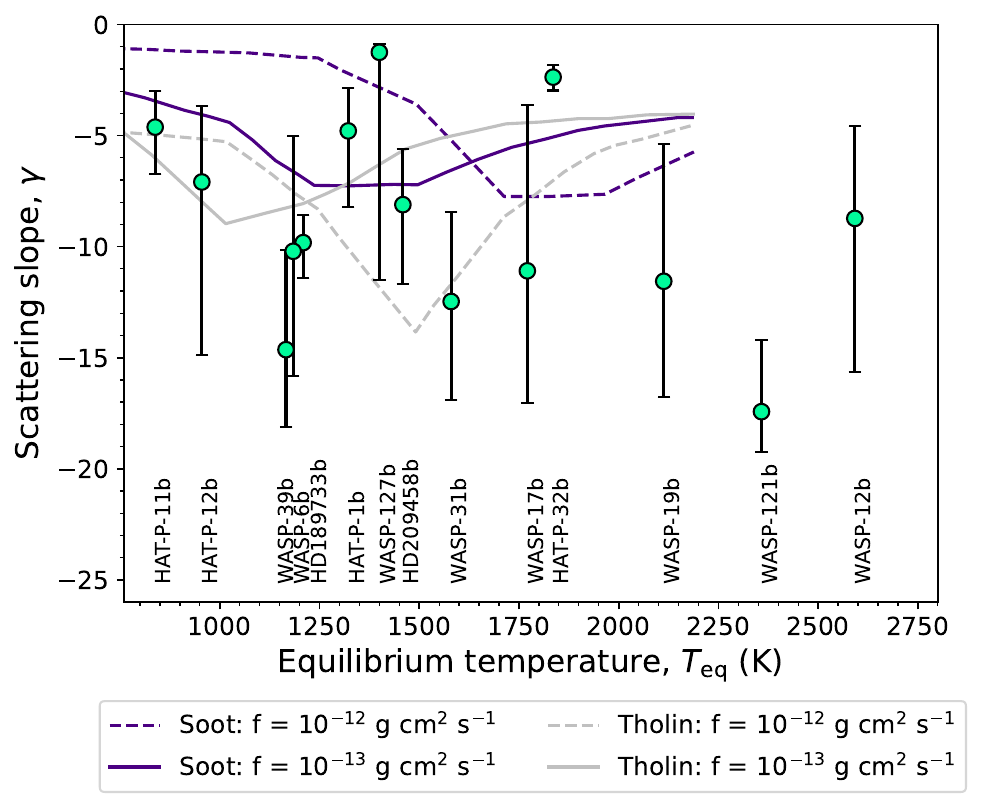}
    \caption{Retrieved scattering slope, $\gamma$, against \teq. The purple and grey lines show the slopes for soot and tholin aerosols, respectively, from \citet{ohno2020ApJ} for two different aerosol mass fluxes ($F$). The models do not extend beyond 2,250\,K and we find no clear trend to any single aerosol profile across the full population of our study.}
    \label{fig:ohno}
\end{figure}

Cloud scattering slopes are generally attributed to small aerosol particles in the upper atmosphere causing Rayleigh-like ($\gamma$\,=\,$-4$) to super-Rayleigh profiles in optical spectra. On average we retrieve scattering slopes 2.2$\times$ Rayleigh for $\lambda_\mathrm{min}$\,=\,0.3\,\microns across all 14 planets in this study. Enhanced Rayleigh-scattering can be attributed to high eddy diffusion coefficients and moderate aerosol mass fractions \citep{ohno2020ApJ}. In Figure~\ref{fig:ohno} we show the profiles for soot and tholin, as a proxy for photochemical hazes, along with our retrieved $\gamma$ values against \teq. The relationship between scattering slope and \teq\ assumes a drag free atmosphere at a pressure of 1\,mbar for particle tracers of 0.01\,\microns based on simulations using different eddy diffusion coefficients \citep{Komacek2019}. We find that the measured scattering cannot be explained by a single aerosol model suggesting that multiple mechanisms (e.g., mineral clouds, \citealt{grant2023}) may also cause enhanced-Rayleigh scattering in exoplanet atmospheres. Further measurements including extended UV or infrared data may help understand the cause of enhanced-scattering in giant exoplanet atmospheres. 

\begin{figure*}
    \centering
    \includegraphics[width = \textwidth]{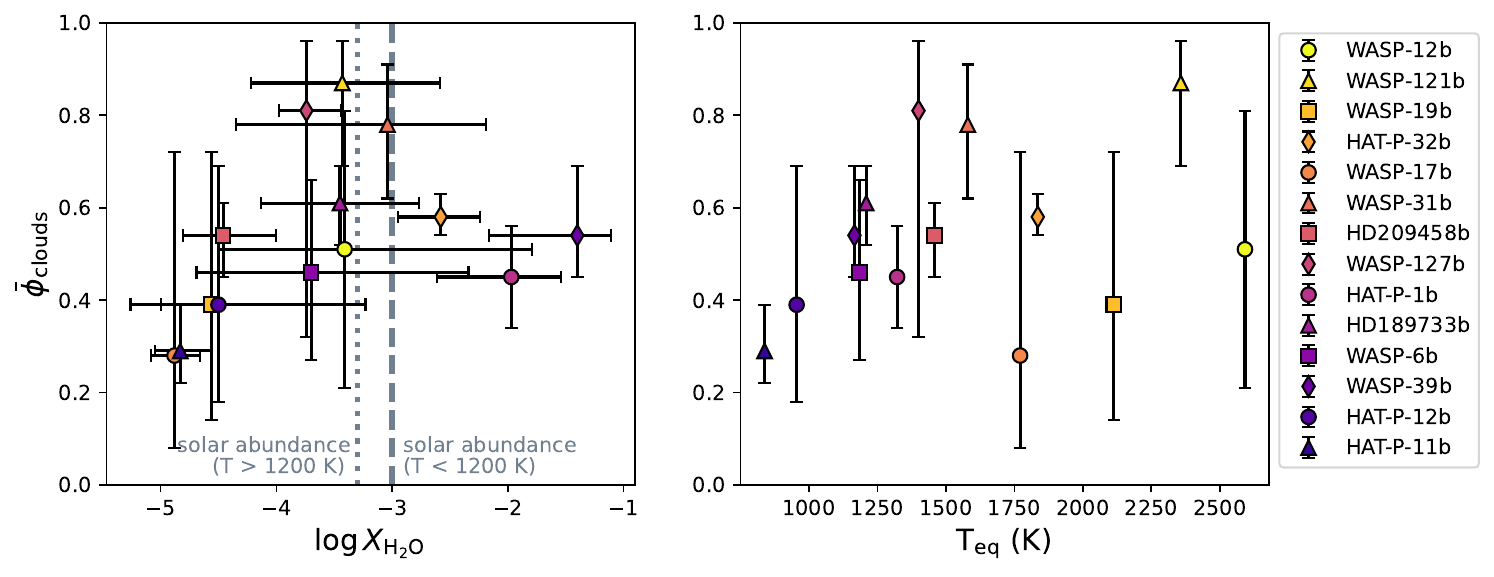}
    \caption{Retrieved cloud fraction, $\bar{\phi}_\mathrm{clouds}$, trends with retrieved water abundance (left) and planetary equilibrium temperature (right). Dashed and dotted lines represent the solar abundance values for temperatures $<$ 1200 K and $>$ 1200 K calculated by \citet{Madhusudhan2012c/o} respectively. The planets are colored by equilibrium temperature from the hottest  (WASP-12b) in yellow to the coldest (HAT-P-11b) in purple.}
    \label{fig:phi_h20_teq}
\end{figure*}
For many planets, clouds are not expected to be present across the entire surface due to changing temperature structures from day-to-night. To account for this, patchy clouds can be invoked to describe the fractional coverage of clouds around the transmitting limb. 
Our retrievals and IC analysis demonstrate that patchy clouds should be considered alongside other cloud parameters when assessing exoplanet transmission spectra. 
We investigate the relationship between patchy clouds, temperature, and water abundance shown in Figure~\ref{fig:phi_h20_teq}.
For planets with water abundances below solar values, a trend of increasing cloud coverage with water abundance could be inferred. 
However, there is only weak evidence to support this from the uncertainties associated with both parameters.
The planets in our sample with the highest water abundance values (WASP-39b, HAT-P-1b and HAT-P-32b) tend towards 50:50 cloud coverage suggesting the potential for morning/evening asymmetries.

We find no apparent trend in changing cloud fraction with equilibrium temperature across the whole range sampled by our 14 planets. We do however note a slight increase in cloud fraction with increasing temperate up to 1600\,K, beyond this uncertainties are too large to determine if there is a turnover or the trend continues to higher temperatures. 
More complex relations which account for a wider range of parameters and micro-physics may be necessary to account for any non-linear trends for example, considering the effect of temperature on the condensation of different species \citet{Gao2020}.
To achieve this, tighter cloud constraints are needed. 
Whilst observationally, higher precision measurements may improve constraints, extending observations to combine transmission and emission spectra (or phase curves) can provide vital spatial information on temperature. 
We also note that extending population investigations to both larger sample sizes and a wider parameter space can help disentangle trends. 

\begin{figure}
    \centering
    \includegraphics[width = 0.48\textwidth]{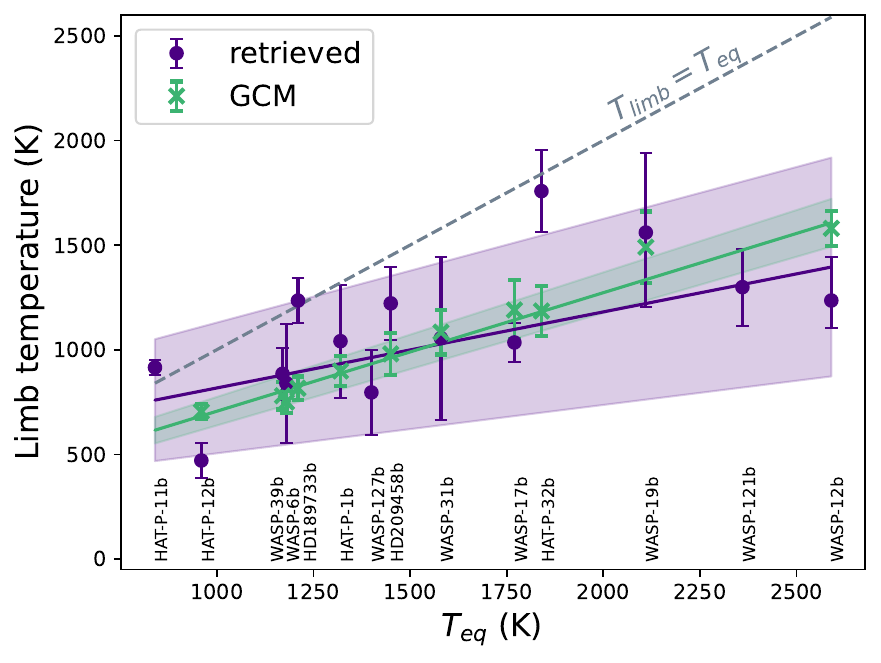}
    \caption{Retrieved limb temperatures plotted against equilibrium temperatures (purple). Plotted in green are GCM limb average temperatures for 11 planets from \cite{Kataria2016}. The gray dashed line marks where limb temperature is equal to planetary equilibrium temperature. Linear fits and $1\sigma$ errors are found for the retrieved limb temperatures and GCM limb temperatures, shown by the purple and green lines and shading respectively.}
    \label{fig:temperatures}
\end{figure}

Through their interaction with internal and external radiation, clouds impact the overall energy budget of a planet and therefore the temperature structure of the atmosphere. 
The formation of clouds along the limbs can be both an indicator and driver of the 3D temperature structure of exoplanet atmospheres. However, we often compare atmospheric thermal profiles to GCM predictions which simulate the radiative convective structure of a planetary atmosphere. 
We evaluate the retrieved temperatures with respect to a population of GCM models presented by \citet{Kataria2016}. Figure \ref{fig:temperatures} presents our retrieved limb temperatures against \teq  alongside average GCM limb temperatures, evaluated across between $10^{-2}$ and $10^{-4}$ bar (representing an estimated range of the observable photosphere) for 11 of the 14 planets. 

Across the retrieved limb temperatures the general trend follows that $T_\mathrm{limb} <$ \teq. Although we include patchy clouds that have been shown to reduce biases towards cool retrieved temperatures that arise from modeling atmospheres with differing compositions between morning and evening terminators \cite{MacDonald2020}, we still find low limb temperatures compared to \teq for the ultra-hot Jupiters WASP-12b and WASP-121b. Our retrieved temperatures reproduce the extreme difference in \teq – T$_\mathrm{limb}$ seen in literature \citep{Kreidberg2015, Evans2018, Pinhas2019, Welbanks2019metallicity} for ultrahot Jupiters where \teq - $T_\mathrm{limb} \sim$ 1,000\,K. This is likely demonstrating where for extreme temperatures, the advective timescale is too large to efficiently transport the vast amount of heat incident on the dayside to the limbs \correction{with a majority re-radiated more efficiently from the atmosphere \citep[e.g.,][]{Fortney2008,komacek2017,Showman2021}}.

However, when comparing the retrieved limb temperatures of the population to GCM temperatures, seven of the 11 planets agree with the GCM temperatures to within $1\sigma$. Fitting a linear relation to $T_\mathrm{limb}-T_\mathrm{eq}$ and $T_\mathrm{GCM}-T_\mathrm{eq}$ gives gradients of 0.36\,$\pm$\,0.13 and 0.57\,$\pm$\,0.03 respectively. The differences in the retrieved and GCM limb temperatures reflect the different aspects of physics they capture. The GCM temperatures are derived from a solar metallicity, cloud free atmosphere, where the divergence from \teq is due to the effects of longitudinal heat transport. In contrast, encoded within our retrieved limb temperatures is the impact of cloud opacity. These two distinct mechanisms have produced similar results within our margin of uncertainty, potentially demonstrating the role of clouds in transport of heat around a planet. Future comparison between modeled and observed limb temperatures may shed more light onto the underlying physical mechanisms driving heat transport in hot Jupiter planets.

\section{Conclusions}\label{sec:conclusions}
Through free-chemistry retrievals with POSEIDON, we set out to explore how retrieved atmospheric parameters inferred from exoplanet transmission spectra improve with the addition of optical data. Initially, the wavelength dependence of retrieved parameters is explored in detail for WASP-39b and HD\,209458b (\S\,\ref{sec:wavelength tests}), to select three spectra ranges to apply to a larger population of planets. We implement retrievals for minimum wavelength ranges, $\lambda_\mathrm{min}$\,=\,0.3, 0.6, and, 1.1\,\microns for a sample of 14 exoplanets with transit spectra from 0.3--4.5\,\microns.
We evaluate the retrieved parameters across the population considering their median values and 1-$\sigma$ uncertainty for each $\lambda_\mathrm{min}$ data range to determine the impact of expanding wavelength range on atmospheric properties. 
To quantify how our knowledge of the atmosphere changes with wavelength range, we implement an information content analysis on the posterior distributions of our retrievals and break down the information content per parameter (\S\,\ref{sec:ICcontent}). 
\correction{As this analysis covers a population of exoplanets, in \S\,\ref{sec:results} we search for trends between atmospheric and planetary parameters.  
We compare our retrieved limb temperatures to temperatures derived from GCM models and investigate the relationship between cloud parameters, water abundance and temperature. More data is required to draw conclusions about population trends.} 

From the spectral range investigation, we find that wavelengths below 0.6\,\microns are necessary to constrain alkali species and cloud scattering parameters $\log{a}$ and $\gamma$, although the scattering slope $\gamma$ is not consistently constrained across the population. This is supported by the information content analysis, where the largest information gains between the $\lambda_{\mathrm{min}}$\,=\,1.1, 0.6 and 0.3\,\microns cutoffs are from the cloud parameters. We find limited impact of the wider optical wavelength coverage on the remaining parameters and abundances. In particular, we note that in general, constraints and median values improve, but not significantly, for cloud pressure level where in most cases, the WFC3/G141 and Spitzer data below 1.1\,\microns provides comparable constraints and retrieved median values to the full 0.3 to 4.5\,\microns spectrum. 

The impact of stellar activity is of particular interest when looking at M-star planets \citep[e.g.,][]{Moran2023}, we show that JWST may not be able to resolve this issue on its own.
HAT-P-11b was an outlier in our planet population due to its location in mass-radius space and that it was the only planet to favor retrievals that account for stellar contamination. For this reason, it provides an important test case for the impact of stellar contamination on retrieved parameters. Without the optical wavelengths, our retrievals do not converge on the same solutions for the stellar parameters as when optical wavelengths are included. With a different stellar spectrum, this leads to vastly different inferences being drawn about the cloud properties and species abundances (particularly water) in the atmosphere. 


Across our population we find a wide range of values for the measured water abundance, in contrast to previous studies which favor low, subsolar abundances, but find no trends with retrieved cloud coverage fractions. We additionally investigate trends of cloud coverage with \teq, finding a tentative suggestion that cloud fraction increases with \teq\ up to 1,600\,K, above which cloud fraction uncertainties become too large to discern significant trends. We show that our retrievals that consider patchy clouds get comparable temperatures to that predicted by GCMs related to \teq. 

Overall this study demonstrates that optical wavelengths below the reach of \jwst are vital to evaluate the cloud properties of exoplanet atmospheres. Studies such as \citet{grant2023} have already demonstrated the utility of optical wavelengths on constraining cloud location and particle size when constraining the composition of clouds via infrared absorption signatures. This supports our analysis which shows that the scattering properties, $\log{a}$ and $\gamma$, are most affected by the inclusion of sub 0.6\,\microns data with improvements in their constraint of over 30\%. 

\vspace{1.5cm}

\correction{The authors thank the anonymous referee for their suggestions and comments.}
We thank L. Alderson for help with defining the idea behind the investigation while writing \hst proposals and wishing something like this existed, N.E. Batalha for helpful comments and discussion around the IC content analysis, and T. Kataria for providing the GCM Pressure-Temperature profiles presented in \citet{Kataria2016}.  
We also thank A. Young and J. Barstow for evaluating the MSc by research thesis in which this work was first presented. 

C.F. is funded by the University of Bristol School of Physics PhD Scholarship Fund. 
H.R.W. was funded by UK Research and Innovation (UKRI) under the UK government’s Horizon Europe funding guarantee for an ERC Starter Grant [grant number EP/Y006313/1]. R.J.M. is supported by NASA through the NASA Hubble Fellowship grant HST-HF2-51513.001, awarded by the Space Telescope Science Institute, which is operated by the Association of Universities for Research in Astronomy, Inc., for NASA, under contract NAS 5-26555.\\

\textit{Data availability:} Supplementary corner plots, spectra and results are available on Zenodo: \dataset[10.5281/zenodo.10407463]{https://doi.org/10.5281/zenodo.10407463}



%


\facilities{HST(STIS), HST(WFC3), Spitzer(IRAC), VLT FORS2, HST(NICMOS)}


\software{numpy \citep{harris2020numpy}, scipy \citep{scipy2020}, matplotlib \citep{hunter2007matplotlib}, POSEIDON \citep{MacDonaldMadhusudhan2017,Macdonald2023}, pymultinest \citep{Buchner2014}.}



\bibliography{paper_bib}{}
\bibliographystyle{aasjournal}

\appendix \label{sec:Appendix}





\section{Planetary and stellar parameters}\label{sec:app:planetary parameters}

\begin{table*}[h!]
    \centering
    \begin{tabular}{ccccccccc}
       Planet  &  $R_* \ (R_\odot)$ & $T_\mathrm{eff} \ (K)$ & [Fe/H] & $\log{g} \ (cgs)$ & $R_p \ (R_J)$ & $M_p \ (M_J)$ & $T_\mathrm{eq} \ (K)$ \\
         \hline \hline
         HAT-P-1 b $^a$ & 1.17 &	5980 &	0.13 &	4.36 &	1.32 &	0.53 &	1320\\
         HAT-P-11 b $^b$ & 0.70 & 4780 & 0.31 &	4.66 &	0.40 &	0.08  &	840\\
         HAT-P-12 b $^c$ & 0.68 & 4670 &	-0.20 &	4.61 &	0.92 &	0.20 &		960 \\
         HAT-P-32 b $^d$ & 1.37 & 6000 &	-0.16 &	4.22 &	1.98 &	0.68 &	1840 \\
         HD\,189733 b & 0.77 $^e$ & 5010 $^f$ &	0.01 $^f$ &	4.49 $^f$ &	1.12 $^e$  &	1.16 $^e$ & 1210 $^e$  \\
         HD\,209458 b & 1.16 $^g$ & 6040 $^f$ &	0.05 $^f$ &	4.30 $^f$ &	1.38 $^g$  &	0.17 $^g$  &	1460 $^g$ \\
         WASP-6 b $^h$ & 0.86 &	5380 &	-0.15 &	4.49 &	1.23 &	0.49 &	1180 \\
         WASP-12 b & 1.66 $^i$ &	6360 $^i$ &	0.33 $^i$ &	4.16 $^i$ &	1.94 $^j$ &	1.47 $^j$ &	2590 $^j$\\
         WASP-17 b &  1.57 $^k$ & 	6490 $^f$  & 0.00 $^f$  & 4.08 $^f$ & 	1.99 $^k$ & 	0.49 $^k$ & 	1770 $^k$ \\
         WASP-19 b & 1.01 $^l$ &	5620 $^l$ &	0.15 $^m$ &	4.42 $^l$ &	1.42 $^l$ &	1.15 $^l$ &	2110 $^l$ \\
         WASP-31 b & 1.25 $^n$ &	6300 $^n$ &	-0.20 $^o$ &	4.76 $^p$ &	1.55 $^n$ &	0.48 $^n$ &	1580 $^q$\\
         WASP-39 b & 0.94 $^c$ &	5460 $^f$ &	-0.01 $^f$ &	4.41 $^f$ &	1.28 $^c$ &	0.28 $^c$ &	1170 $^c$\\
         WASP-121 b & 1.46 $^r$ & 6340 $^f$ &	0.24 $^f$ &	4.17 $^f$ &	1.75 $^s$&	1.16 $^s$ &  2360 $^r$ \\
         WASP-127 b & 1.33 $^t$ & 5850 $^f$ &	-0.16 $^f$ &	4.22 $^f$ &	1.31 $^t$ &	0.16 $^t$ &	1400 $^u$\\
    \end{tabular}
    \caption{Stellar and planetary parameters for the sample of 14 exoplanets used in this investigation, where values have been obtained from the following studies: \cite{Nikolov2014}$^a$; \cite{Southworth2011hat11}$^b$; \cite{Mancini2018}$^c$; \cite{Wang2019}$^d$; \cite{Addison2019}$^e$; \cite{Polanski2019}$^f$; \cite{Southworth2010} $^g$; \cite{TregolanReed2015} $^h$; \cite{Collins2017} $^i$; \cite{Chakrabarty2019} $^j$; \cite{Anderson2011w17} $^k$; \cite{cortez2020} $^l$; \cite{Knutson2014} $^m$; \cite{Anderson2011w31} $^n$; \cite{Bonomo2017} $^o$; \cite{Mortier2013} $^p$; \cite{Sing2016}$^q$; \cite{Delrez2016} $^r$; \cite{Bourrier2020} $^s$; \cite{Seidel2020} $^t$; \cite{Lam2017} $^u$. }
    \label{tab:planet parameters}
\end{table*}





\end{document}